\documentclass[iop,twocolumn]{emulateapj}

\usepackage{placeins}
\usepackage{graphics}
\usepackage{color} 
\usepackage{amsmath}

\def\2pr{^{\prime \prime}} 
\def\deg{^{\circ}} 
\def\greatsim{\mathrel{\raise.3ex\hbox{$>$\kern-.75em\lower1ex\hbox{$\sim$}}}} 
\def\lesssim{\mathrel{\raise.3ex\hbox{$<$\kern-.75em\lower1ex\hbox{$\sim$}}}} 
\def\gs{\mathrel{\raise0.27ex\hbox{$>$}\kern-0.70em 
\lower0.71ex\hbox{{$\scriptstyle \sim$}}}} 
\def\ls{\mathrel{\raise0.27ex\hbox{$<$}\kern-0.70em 
\lower0.71ex\hbox{{$\scriptstyle \sim$}}}}

\slugcomment{Submitted to {\em Astrophysical Journal}} 
 
\shorttitle{BOSS Composite Quasar Spectra} 
\shortauthors{Jensen et al.}

\begin{document} 
 
\title{Spectral Evolution in High Redshift Quasars from the Final Baryon Oscillation Spectroscopic Survey Sample} 
 
\author{
Trey~W.~Jensen\altaffilmark{1}, 
M.~Vivek\altaffilmark{1},
Kyle~S.~Dawson\altaffilmark{1},
Scott~F.~Anderson\altaffilmark{2},
Julian~Bautista\altaffilmark{1},
Dmitry~Bizyaev\altaffilmark{3,4},
William~N.~Brandt\altaffilmark{5},
Joel~R.~Brownstein\altaffilmark{1},
Paul~Green\altaffilmark{6}, 
David~W.~Harris\altaffilmark{1}$^{,}$\altaffilmark{7}, 
Vikrant~Kamble\altaffilmark{1},
Ian~D.~McGreer\altaffilmark{8},
Andrea~Merloni\altaffilmark{9},
Adam~Myers\altaffilmark{10},
Daniel~Oravetz\altaffilmark{3},
Kaike~Pan\altaffilmark{3},
Isabelle~P\^aris\altaffilmark{11},
Donald~P.~Schneider\altaffilmark{5}$^{,}$\altaffilmark{12},
Audrey~Simmons\altaffilmark{3},
Nao~Suzuki\altaffilmark{13}
}
\altaffiltext{1}{ 
Department of Physics and Astronomy,  
University of Utah, Salt Lake City, UT 84112, USA 
}
\altaffiltext{2}{
Department of Astronomy,
University of Washington, Seattle, WA 98195, USA}

\altaffiltext{3}{Apache Point Observatory and New Mexico State
University, P.O. Box 59, Sunspot, NM, 88349-0059, USA}

\altaffiltext{4}{Sternberg Astronomical Institute, Moscow State
University, Moscow}

\altaffiltext{5}{
Institute for Gravitation and the Cosmos,
Pennsylvania State University, University Park, PA 16802, USA}

\altaffiltext{6}{
Harvard-Smithsonian Center for Astrophysics,
Cambridge, MA 02138, USA}

\altaffiltext{7}{
Department of Science,
United States Coast Guard Academy, New London, CT 06320, USA}

\altaffiltext{8}{
Steward Observatory,
University of Arizona, Tucson, AZ 85721, USA}

\altaffiltext{9}{
Max-Planck-Institut f\"{u}r Extraterrestrische Physik,
Giessenbachstra{\ss}e, 85748 Garching, Germany}

\altaffiltext{10}{
Department of Physics and Astronomy,
University of Wyoming, Laramie, WY 82071, USA}

\altaffiltext{11}{
LAM (Laboratoire d'Astrophysique de Marseille),
Aix Marseille Universit\'{e}, UMR 7326, 13388, Marseille, France}

\altaffiltext{12}{
Department of Astronomy and Astrophysics,
The Pennsylvania State University, University Park, PA 16802}

\altaffiltext{13}{ 
Kavli Institute for the Physics and Mathematics of the Universe (IPMU), The University of Tokyo, Kashiwanoha 5-1-5, Kashiwashi, Chiba, Japan}

\date{\today} 

\email{tjens150@gmail.com}

\begin{abstract}
We report on the diversity in quasar spectra from the Baryon
Oscillation Spectroscopic Survey.
After filtering the spectra to mitigate selection effects and Malmquist bias associated with a nearly flux-limited sample, we create high signal-to-noise ratio composite spectra from 58,656 quasars ($2.1 \leq z \leq 3.5$), binned by luminosity, spectral index, and redshift.
With these composite spectra, we confirm the traditional Baldwin effect (i.e. the anti-correlation of \textsc{C iv} equivalent width and luminosity) that follows the relation $W_{\lambda} \propto L^{\beta_{w}}$ with slope $\beta_w = -0.35 \pm 0.004$, $-0.35 \pm 0.005$, and $-0.41 \pm 0.005$ for $z = 2.25, 2.46, \textrm{ and } 2.84$, respectively. 
In addition to the redshift evolution in the slope of the Baldwin effect, we find redshift evolution in average quasar spectral features at fixed luminosity. The spectroscopic signature of the redshift evolution is correlated at 98\% with the signature of varying luminosity, indicating that they arise from the same physical mechanism.
At a fixed luminosity, the average \textsc{C iv} FWHM decreases with increasing redshift and is anti-correlated with \textsc{C iv} equivalent width. The spectroscopic signature associated with \textsc{C iv} FWHM suggests that the trends in luminosity and redshift are likely caused by a superposition of effects that are related to black hole mass and Eddington ratio.
The redshift evolution is the consequence of a changing balance between these two quantities as quasars evolve toward a population with lower typical accretion rates at a given black hole mass.
\end{abstract}

\section{\textbf{Introduction}}\label{sec:intro}

Spectral properties of quasars depend on the physical conditions of the accreting black hole and
on the geometry of the emitting regions.  Quasars display striking similarities such as broad emission lines
and continua that follow an approximate powerlaw relationship in the near UV and optical wavelengths \citep[][and references therein]{merloni16a}.
However, the exact properties of the continuum and emission lines vary across the population of
quasars in a manner that is poorly understood \citep[][for instance]{richards11,shen14,baskin15a}.
Our understanding of quasar diversity has improved in recent years through applications of several analysis techniques
to the sample of spectroscopically-observed quasars. Among these techniques, the identification of trends
among observed parameters across the whole population of quasars, principal component analysis \citep[PCA;][for instance]{boroson92},
and the use of composite spectra are among the most commonly employed.  These techniques
have allowed characterization of the most significant correlations between the continuum and emission line properties of quasars.
Constraining the trends across the sample of quasars allows us to better understand the underlying physics of
these systems and informs the modeling of quasar spectra for better redshift determination in studies of large-scale clustering.

The application of PCA decomposition to large samples of quasars allows the
empirical characterization of the most significant correlations between spectroscopic properties.
The PCA technique builds a minimum set of basis vectors across the observed parameter space to capture
the total variance in the sample. This method has revealed a striking relationship in which the relative optical to X-ray
flux and the equivalent width (EW) of Fe \textsc{ii} are shown to be directly correlated with the EW of [O \textsc{iii}] and the
full width at half maximum (FWHM) of H$_{\beta}$.  Known as Eigenvector 1, this trend was first identified by
\citet{boroson92} and later confirmed in additional samples \citep{shang03,yip04,sulentic09}. 
The Eddington ratio is believed to be the physical driver of the Eigenvector 1 correlations \citep[e.g.,][]{boroson02,marziani03,shen14}. 

The similarity of quasar spectra over wide ranges of luminosity and redshift has enabled the use of composite spectra
to study their global properties.  Composite spectra offer several advantages over the other methods by virtue of their high
signal-to-noise ratios and the tendency to average out the intrinsic peculiarities of any single source. 
Weaker emission lines unidentifiable in single objects are detectable in composite spectra. 
This method has been employed over a wide range of quasar samples such as the Large Bright Quasar Survey \citep[LBQS;][]{francis91a, hewett95a}, 
FIRST Bright Quasar Survey \citep[FBQS;][]{white00b,brotherton01a}, Sloan Digital Sky Survey \citep[SDSS;][]{york00a,vandenberk01a},
Hubble Space Telescope \citep[HST;][]{zheng97,telfer02}, HST- Far Ultraviolet Spectroscopic Explorer \citep[FUSE;][]{scott04a},
HST- Cosmic Origin Spectrograph \citep[COS;][]{shull12,stevans14a,tilton16a}, Baryon Acoustic Oscillation Spectroscopic Survey \citep[BOSS;][]{dawson13a,harris16a},
and VLT-XSHOOTER \citep{selsing16}.

Despite the general similarities, quasars differ in their luminosity, spectral index,
and emission line features.  Unrestricted composite spectra mask the intrinsic dispersion in these key parameters
which would otherwise provide clues to underlying physics. Composite spectra constructed from a subsample of sources binned
by common observed properties allow the study of correlated properties in a manner which does not necessarily
impose the orthogonality constraints of the PCA approach. For example, \citet{richards11} constructed composite spectra from $\approx$30,000 quasars from the 7th Data Release of SDSS as a function of
\textsc{C iv} equivalent width and blueshift to reveal the empirical relationships between emission features of line species and the continuum.
They reported that the \textsc{C iv} emission parameters capture an important relationship between the disk and wind components of the
broad emission lines that appears to be sensitive to the shape of the ionizing continuum.
\citet{ivashchenko14} constructed composite spectra binned by spectral index and showed that their composite spectra
resulted in higher average systemic redshift estimates than the SDSS templates.

One of the most prominent observed relations in quasar spectra is the anti-correlation between the rest-frame equivalent width of UV emission
(particularly \textsc{C iv}) and continuum luminosity, first identified by \citet{vbaldwin77}.  
The dependence of equivalent widths on luminosity is commonly referenced as the Baldwin effect (BE).  The BE is now well established for
nearly all broad emission lines. \cite{kinney90a} and \cite{zamorani92a} expanded the original work to larger ranges of quasar luminosity and redshift,
thus revealing similar effects in Mg \textsc{ii} \citep[see][for an outline]{sulentic00a}.  \cite{green01a} discovered a significant BE in UV iron emission, although
also found that UV iron emission may have stronger correlation with redshift, implying an evolutionary effect rather than a direct
dependence on luminosity.
\cite{croom02a} and \cite{dietrich02a} revealed the effect in narrow emission lines using composite quasar methods. 
The BE in the X-ray Fe K$\alpha$ feature has also been identified in a number of studies \citep[e.g.,][]{page05,wu09a,ricci13}. 
\citet{vbaldwin89} and \citet{dietrich02a} showed that the slope of the BE steepens with increasing ionization potential.

In quasars with multiple observations, the intrinsic BE associated with change in luminosity
has also been identified for different broad emission lines.
The same trend of a decreasing EW with increasing luminosity is seen, but with a steeper slope than the
ensemble BE that is observed over a population of quasars \citep[see][]{pogge92a,gilbert03,cackett06}.
Among the several physical explanations proposed to account for the BE and its ionization dependence,
the most popular is a prediction that the continuum softens at short wavelengths with increasing luminosity.
Models predicting this relationship result in fewer ionizing photons in luminous quasars \citep{netzer92a,korista98a}. 

Despite the myriad of studies and observations of the BE, there are no rigorous models that provide a complete explanation
of the effect's origin \citep[for a review, see ][and references therein]{shields07a}. The BE remains an empirical correlation,
so it is not clear if the variations in line strengths are driven by luminosity or by a more fundamental parameter such as
black hole mass, Eddington ratio, or orientation.  Because these parameters are all correlated with luminosity,
it is difficult to disentangle the origins of the effect \citep{bachev04a,shields07a,bian12a}. 
By creating composite spectra over different ranges in FWHM of \textsc{C iv} emission, \citet{warner04} provide evidence that
black hole mass may be largely responsible for the EW variations associated with luminosity.
In contrast, after binning on Eddington ratio, the composite spectra do not show BE-like behavior. 
However, \citet{kollmeier06a} suggest that the interpretation of these results may be clouded by residual selection effects.

In this study, we explore spectral diversity using composite spectra over an unprecedented sample of 112,000 quasars.
{ We mitigate biases in the distribution of colors and bolometric luminosities that are intrinsic to a flux-limited survey in each of the three redshift intervals.   We generate high signal-to-noise ratio composite spectra at each redshift in bins of constant luminosity and spectral index.}

We use these composite spectra to study the changes in quasar spectral features as a function of redshift, luminosity,
and spectral index.  By characterizing diversity through a linear interpolation between composite spectra, this approach
is similar to PCA.  Unlike the PCA approach, we impose no orthogonality restriction which enables one to find common modes in
diversity associated with each of these three parameters.

The paper is structured as follows: Section~2 describes the data, the binning and sampling methods, and the construction of composite spectra.
Section~3 presents the variation in spectral signature associated with luminosity, spectral index, and redshift.
In Section~4, we probe different observed quasar parameters that may be responsible for the variation in spectral signatures seen in luminosity and redshift.
Section~5 summarizes our findings and discusses possible future work.
Throughout, we assume a $\Lambda CDM$ cosmology with $H_{0} = 70\, {\rm km}\, {\rm s}^{-1}$, $\Omega_{M} = 0.3$, and $\Omega_{\Lambda} = 0.7$.

\section{\textbf{Data}}\label{sec:data}
We utilize quasars observed in the Baryon Oscillation Spectroscopic Survey \citep[BOSS;][]{dawson13a}, a component of SDSS-III \citep{eisenstein11a}. 
Using the Apache Point 2.5-meter Telescope \citep{gunn06a}, SDSS-III and BOSS concluded in 2014. The final public data release is the 12th in the series from SDSS, comprising 2,401,952 galaxy and 477,161 quasar spectra \citep[DR12;][]{alam15a}. 

BOSS was designed to characterize the Baryonic Acoustic Oscillation (BAO) feature first detected roughly a decade ago in the clustering of galaxies \citep{eisenstein05a,cole05a}. 
BOSS used luminous red galaxies and the Lyman-$\alpha$ forest observed in the absorption of quasar spectra
to measure the BAO signal in the correlation of matter. 
Spectroscopic observations are performed using plates with a radius of 1.5$\deg$ on the sky \citep{smee13a}.
On each plate, there are 1000 holes drilled for 2 arcsecond diameter fibers, where approximately 600-700 fibers
are used for galaxy targets and 160-200 fibers are used for quasar targets. A total of 20 fibers are reserved for standard
stars with $ugriz$ colors consistent with main sequence F stars.
The most recent BOSS BAO measurements from galaxies are reported in \citet{alam16a} and from the Lyman-$\alpha$ forest in \citet{delubac15a}.

The majority of quasar targets were selected from SDSS $ugriz$ imaging data \citep{fukugita96a}. Additional selections included imaging data from the Galaxy Evolution Explorer \citep[GALEX;][]{martin05a} and UKIRT Infrared Deep Sky Survey \citep[UKIDSS;][]{lawrence07a}. Together, the selection methods \citep[][\citeyear{bovy12a}]{richards09a, yeche10a,kirkpatrick11a,bovy11a} resulted in a total density of confirmed Lyman-$\alpha$ quasars exceeding 15 deg$^{-2}$ \citep{ross12a}. The quasars used in \citet{delubac15a} span the redshift range $2.1 \leq z \leq 3.5$. In the analysis presented in this work, we define a control sample based on spectroscopically observed properties. We are therefore less susceptible to target selection biases in the BOSS sample and we include all targets regardless of their selection algorithm.

\subsection{Data Reduction}\label{subsec:sdssboss}

Once the observations are completed, there are two major steps that comprise data reduction. One-dimensional spectra are extracted and calibrated from the raw, two-dimensional CCD images \citep{stoughton02a}. The extracted spectra are flux-calibrated based on model spectra fit to the aforementioned F stars found on each plate. The second step of the data reduction pipeline performs a classification and a redshift estimate for each spectrum \citep{bolton12a}.

Significant changes have been made to the spectral extraction and coaddition of individual exposures since the DR12 release. These changes are motivated by the challenges of classifying lower signal-to-noise ratio spectra found in the program succeeding BOSS. Like BOSS, the extended Baryon Oscillation Spectroscopic Survey \citep[eBOSS;][]{dawson16a} is the cosmological component of an iteration of the SDSS; this time SDSS-IV (Blanton et al. in prep.). The first major change corrects a known bias in the coaddition of individual exposures. This correction has significant impact on the classification of galaxy spectra and is described in \citet{hutchinson16a}. The second change pertains to flux calibration and directly impacts the interpretation of quasar spectra. Because it is relevant to this work, we provide a description below.

As described in \citet{dawson13a}, the atmospheric differential refraction (ADR) bears a different signature in quasar spectra than the standard stars due to a choice in the placement of fibers where quasars are drilled to maximize the flux at 4000 \AA\ instead of 5400 \AA\ for standard stars. \citet{harris16a} provide an empirical, global correction to this effect by using standard stars observed in the same fashion as quasar targets. Likewise, \citet{margala15a} derive a first principles estimate of ADR for each quasar target and apply corrections to the final coadded spectra. In the newest reductions, we have improved upon both of these analyses { by applying the flux corrections derived from our ADR model to quasar spectra} in individual exposures before any flux calibration is applied. This approach has the distinct advantage of correcting for ADR flux loss in 15-minute increments corresponding to the individual exposure time, rather than making a correction to the final coadded spectrum based on the averaged flux loss.

\begin{figure}[htb!]
\centering
\includegraphics[scale=0.42,trim=0cm 0cm 0cm .95cm,clip]{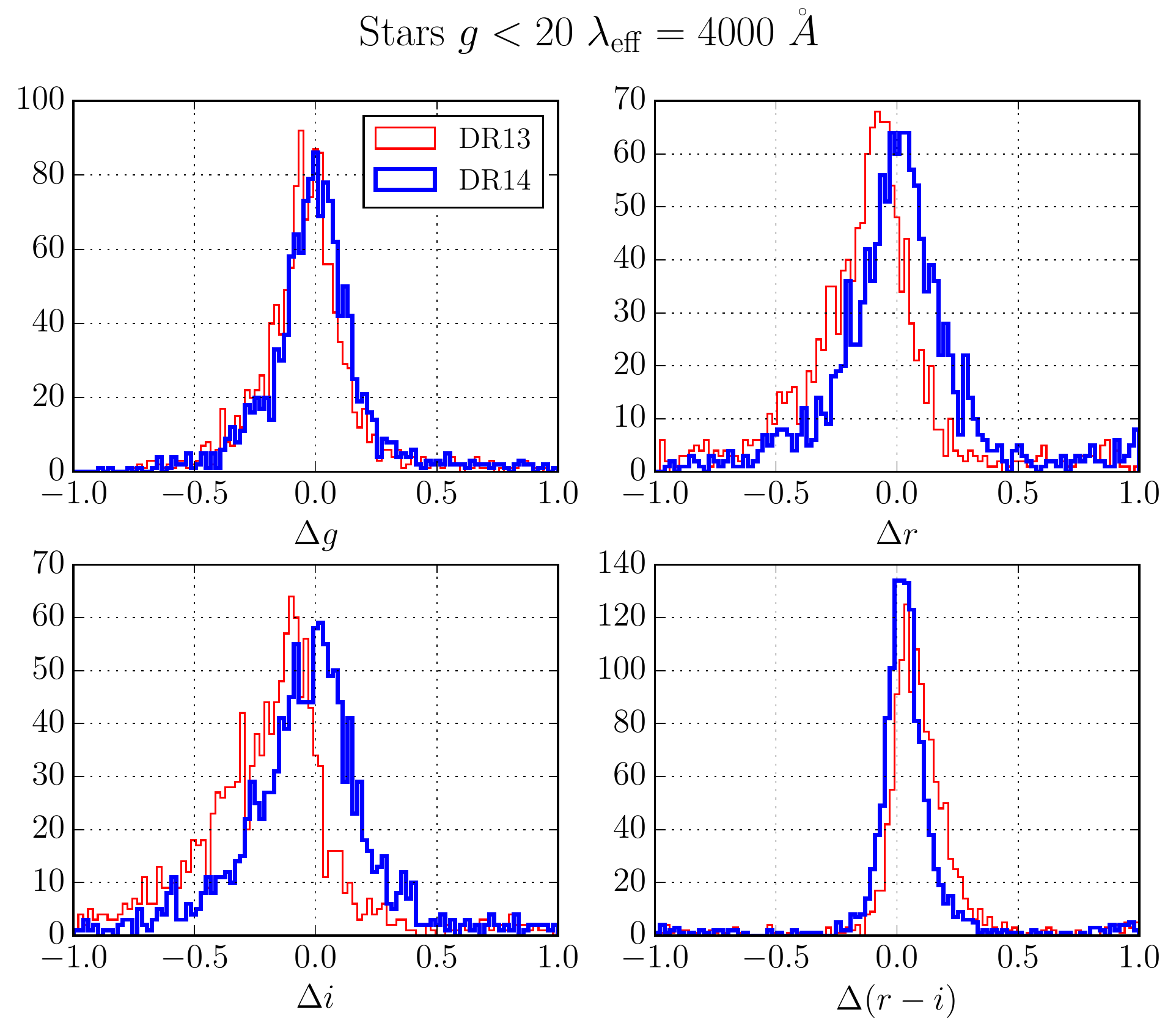}
\caption{Residual bias between SDSS imaging and BOSS spectrophotometry computed for 1330 stars brighter than $g = 20$ that were targeted as quasars. The histogram for each panel shows the difference between the DR13 reductions (red) and the new improved DR14 reductions (blue) with ADR corrections. Residuals are presented for $gri$ and $r-i$.}
\label{fig:newpipe}
\end{figure}

The new data reduction pipeline is complete and has been used to reprocess all quasar data taken during BOSS. Tagged as {\tt v\_5\_10}, these reductions of BOSS data and all eBOSS data taken in 2014--2016 will be released as DR14 in Summer 2017. The results show significant improvement in the flux calibration of quasar spectra. We quantify spectrophotometric errors by examining the residuals in the spectra of stars that appear as contaminants in the selection algorithms for BOSS quasars because these were observed in the same manner as quasars. We compute synthetic magnitudes in filters corresponding to $gri$ from these stellar spectra using the DR13 reductions and the new DR14 reductions. By comparing those synthetic magnitudes to the $gri$ magnitudes from the SDSS imaging data, we can characterize the relative systematic errors in flux calibration. As shown in Figure~\ref{fig:newpipe}, the residuals in $gri$ magnitudes and $r-i$ color decrease significantly between the DR13 reductions with no ADR corrections and the new improved DR14 reductions with ADR corrections. The median residual in ($gri$, $r-i$) decreases from ($-0.041$, $-0.109$, $-0.177$, 0.078) to ($-0.001$, 0.004, $-0.022$, 0.032).

For our analysis, we use the new calibrated spectra found in DR14 and the classifications from the quasar catalog released along with DR12 \citep[DR12Q;][]{paris16a}.
Several techniques were implemented in DR12Q to improve the redshift precision, classification, and identification of peculiar quasars relative to the BOSS pipeline classifications. We use the PCA technique as the estimate of the redshift. 
 
\subsection{Sample Selection}\label{subsec:sample}


There are 175,294 quasars in the redshift range $2.1 \leq z \leq 3.5$ included in the DR12Q catalog. We restrict the population to that redshift range as it has the most complete selection and will allow better understanding of the quasar population used in BOSS Lyman-$\alpha$ forest studies. We remove broad absorption line quasars (BALs) that are identified in DR12Q with the flag {\tt BAL\_FLAG\_VI} due to their influence on emission lines. Damped Lyman-$\alpha$ (DLA) quasars identified by the same technique as those in the catalog from \citet{noterdaeme12a} are removed to suppress contamination of quasar continua from metal absorption lines. After application of these constraints, 143,341 quasar spectra remain. Additionally, all quasars observed at an airmass above 1.2 are removed, as the spectrophotometric correction \citep[see][]{harris16a} may be less certain at high airmass.
 The remaining sample is large enough that removing the $\sim$30,000 high airmass quasars will not significantly affect the final results.  In total, 112,000 quasars were included in this study. The sample size at each phase of filtering is reported in Table~\ref{tab:sampletable}.
\begin{table}[b!]
\centering
\caption{The number of quasars remaining in the sample after each sample selection criterion is applied.}
\begin{tabular}{c r}
\hline\hline
Criterion & Remaining Quasars \\
\hline
   $2.1 \leq z \leq 3.5$ & 175,294 \\
   Remove BAL, DLA & 143,341 \\
   Airmass $\leq$ 1.2 & 112,000 \\
   Final cut after reshape & 58,656 \\
   \hline
\end{tabular}
\label{tab:sampletable}
\end{table}

{ In any flux-limited sample, the minimum luminosity increases with increasing redshift, so the range of allowed black hole masses and Eddington ratios able to account for a given luminosity at a higher redshift is then narrowed relative to a lower redshift \citep[e.g., Fig. 2 of][AAp, 570, A96]{sulentic14a}. To inspect spectral variations, it is important to choose parameters that can be derived from both the low and high redshift ranges of the sample in identical fashion.  While it is difficult to directly measure black hole mass and Eddington ratio, luminosity and spectral index impose selection effects and can be easily measured.}

In this work we explore spectral diversity against three parameters: bolometric luminosity ($L_{bol}$), spectral index ($\alpha_{\lambda}$), and redshift ($z$). These are all direct spectroscopic observables; $L_{bol}$ is the parameter which is often used to explain variations in emission line properties, while $\alpha_{\lambda}$ plays a major role in determining quasar colors. Because color measurements are key to the selection of quasar targets, variations in the spectral index with variations in emission line strengths can cause different degrees of completeness in different redshift ranges \citep[e.g.,][]{richards02a}.

The quantity $L_{bol}$ is calculated from the monochromatic luminosity at 1350 \AA, where $L_{bol} = c \times L_{1350}$, $L_{1350}$ is the continuum luminosity at 1350 \AA, and $c = 4$, consistent with the values in \citet{xuy08a} and \citet{richards09a}.  The spectral index is defined by fitting the powerlaw $f(\lambda) = b \lambda^{\alpha_{\lambda}}$ to the estimated quasar continuum. We choose a range of wavelengths to model the quasar continuum in which visual inspection of the high signal-to-noise ratio composite spectrum from \citet{harris16a} reveals no clear features. We use two regions: 1440 \AA\ to 1480 \AA\ and 2160 \AA\ to 2230 \AA. We adopt FWHM measurements from DR12Q, and record $L_{bol}$ and $\alpha_{\lambda}$ and their associated measurement errors for every one of the 112,000 quasars. 


\subsection{Composite Spectra with Controlled Selection}\label{subsec:binning}


\begin{figure}[htb!]
\centering
\includegraphics[scale=0.46,trim=1.25cm 12.6cm 1cm 1.8cm,clip]{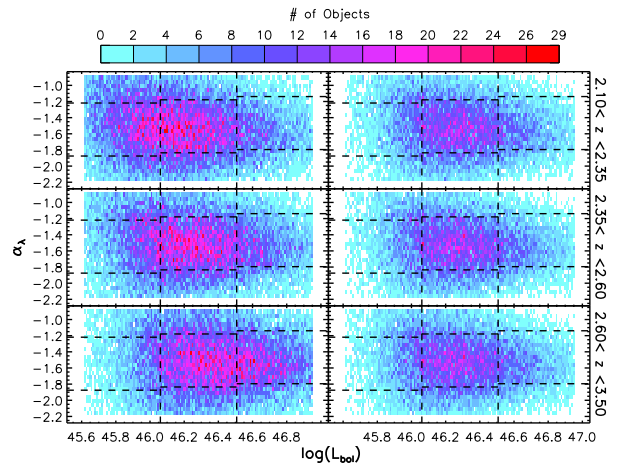}
\caption{$\alpha_{\lambda}$ versus log($L_{bol}$) two-dimensional histogram density plot. {\bf Left:} The native distributions of $\textrm{log}(L_{bol})$ and $\alpha_{\lambda}$ in three panels corresponding to each redshift bin in ascending order from top to bottom. {\bf Right:} The distributions after modification plotted in the same fashion. The dashed lines signify where we cut the distributions, ultimately forming 27 bins.}
\label{fig:histmolds}
\end{figure}

We divide the sample of 112,000 quasars according to distinct observable properties to generate a series of composite spectra. From these composite spectra, we explore the systematic changes in spectral features associated with changes in luminosity, spectral index, and redshift.

Because the quasar sample used in our analysis spans a large redshift range, and is selected primarily by $ugriz$ colors, the distribution of $L_{bol}$ and $\alpha_{\lambda}$ is subject to artificially-induced redshift dependence. Malmquist bias introduces a selection effect in which lower luminosity quasars are preferentially selected at low redshift and are typically rejected at high redshift. Likewise, the color selection performed in the observer frame samples quasars with different spectral indices at different redshifts due to the presence of emission lines. When generating composite spectra, we must control diversity within each bin such that there is no redshift evolution in $L_{bol}$ or $\alpha_{\lambda}$.

We first bin over three redshift intervals from $2.1 \le z \le 2.35$, $2.35 \le z \le 2.6$, and $2.6 \le z \le 3.5$, chosen to ensure a similar number of objects in each bin. We next require the distribution of $L_{bol}$ and $\alpha_{\lambda}$ in each of these three redshift bins to have the same high-order statistics by first constructing a two-dimensional histogram spanning these two parameters. We clip the distributions so that only objects within two standard deviations from the mean are included.
{ At each bin ($L_{bol}$, $\alpha_{\lambda}$) in the histogram, we identify the redshift interval with the minimum number of objects and randomly down-sample from the other two redshift intervals until all three have the same number of objects.}

\begin{table*}[!htb]
\centering
\caption{The median value of all single measurements in each composite spectrum bin for redshift, $\textrm{log}(L_{bol})$, $\alpha_{\lambda}$, signal-to-noise ratio (S/N), and \textsc{C iv} FWHM, and the number of objects in each bin. \textsc{C iv} FWHM are obtained from the DR12Q catalog.}
\begin{tabular}{c c c c r c c}
\hline\hline
Bin & $z$ & $\textrm{log}(L_{bol})$ & $\alpha$ & \# of Spectra & \textsc{C iv} FWHM ($\textrm{km s}^{-1}$) & S/N \\
\hline

$(z_{\rm low},L_{\rm low},\alpha_{\rm low})$ & 2.25 & 45.94 & -2.06 & 843 & 4118 & 2.40 \\
$(z_{\rm low},L_{\rm low},\alpha_{\rm mid})$ & 2.25 & 45.95 & -1.51 & 2531 & 4137 & 2.48 \\
$(z_{\rm low},L_{\rm low},\alpha_{\rm hi})$ & 2.26 & 45.91 & -0.98 & 1602 & 4242 & 2.34 \\
$(z_{\rm low},L_{\rm mid},\alpha_{\rm low})$ & 2.25 & 46.26 & -1.98 & 1936 & 4521 & 4.86 \\
$(z_{\rm low},L_{\rm mid},\alpha_{\rm mid})$ & 2.25 & 46.28 & -1.52 & 6735 & 4518 & 5.15 \\
$(z_{\rm low},L_{\rm mid},\alpha_{\rm hi})$ & 2.25 & 46.24 & -1.03 & 1630 & 4574 & 4.95 \\
$(z_{\rm low},L_{\rm hi},\alpha_{\rm low})$ & 2.26 & 46.64 & -1.91 & 843 & 4982 & 9.99 \\
$(z_{\rm low},L_{\rm hi},\alpha_{\rm mid})$ & 2.25 & 46.66 & -1.52 & 3100 & 4865 & 10.64 \\
$(z_{\rm low},L_{\rm hi},\alpha_{\rm hi})$ & 2.24 & 46.62 & -1.05 & 332 & 4833 & 10.22 \\
$(z_{\rm mid},L_{\rm low},\alpha_{\rm low})$ & 2.45 & 45.94 & -2.06 & 838 & 3985 & 2.15 \\
$(z_{\rm mid},L_{\rm low},\alpha_{\rm mid})$ & 2.44 & 45.95 & -1.51 & 2536 & 3968 & 2.19 \\
$(z_{\rm mid},L_{\rm low},\alpha_{\rm hi})$ & 2.44 & 45.91 & -0.97 & 1594 & 4075 & 2.05 \\
$(z_{\rm mid},L_{\rm mid},\alpha_{\rm low})$ & 2.45 & 46.26 & -1.98 & 1942 & 4423 & 4.32 \\
$(z_{\rm mid},L_{\rm mid},\alpha_{\rm mid})$ & 2.46 & 46.28 & -1.52 & 6739 & 4423 & 4.56 \\
$(z_{\rm mid},L_{\rm mid},\alpha_{\rm hi})$ & 2.46 & 46.25 & -1.03 & 1622 & 4417 & 4.36 \\
$(z_{\rm mid},L_{\rm hi},\alpha_{\rm low})$ & 2.46 & 46.64 & -1.91 & 828 & 4932 & 9.30 \\
$(z_{\rm mid},L_{\rm hi},\alpha_{\rm mid})$ & 2.47 & 46.66 & -1.52 & 3124 & 4869 & 9.56 \\
$(z_{\rm mid},L_{\rm hi},\alpha_{\rm hi})$ & 2.47 & 46.62 & -1.04 & 329 & 4805 & 9.17 \\
$(z_{\rm hi},L_{\rm low},\alpha_{\rm low})$ & 2.78 & 45.94 & -2.07 & 823 & 3634 & 1.66 \\
$(z_{\rm hi},L_{\rm low},\alpha_{\rm mid})$ & 2.76 & 45.95 & -1.52 & 2545 & 3479 & 1.74 \\
$(z_{\rm hi},L_{\rm low},\alpha_{\rm hi})$ & 2.78 & 45.92 & -0.98 & 1608 & 3470 & 1.60 \\
$(z_{\rm hi},L_{\rm mid},\alpha_{\rm low})$ & 2.85 & 46.26 & -1.98 & 1944 & 4058 & 3.24 \\
$(z_{\rm hi},L_{\rm mid},\alpha_{\rm mid})$ & 2.84 & 46.28 & -1.52 & 6733 & 4019 & 3.44 \\
$(z_{\rm hi},L_{\rm mid},\alpha_{\rm hi})$ & 2.89 & 46.25 & -1.03 & 1631 & 3987 & 3.15 \\
$(z_{\rm hi},L_{\rm hi},\alpha_{\rm low})$ & 2.93 & 46.65 & -1.92 & 825 & 4664 & 6.71 \\
$(z_{\rm hi},L_{\rm hi},\alpha_{\rm mid})$ & 2.91 & 46.66 & -1.52 & 3106 & 4586 & 7.13 \\
$(z_{\rm hi},L_{\rm hi},\alpha_{\rm hi})$ & 2.95 & 46.62 & -1.04 & 337 & 4469 & 6.45 \\

\hline
\end{tabular}
\label{tab:bintable}
\end{table*}

The reshaping of the quasar population results in the removal of $\sim 45\%$ of the spectra (see Table~\ref{tab:sampletable}). We divide each of the three redshift bins into thirds of $\textrm{log}(L_{bol})$, denoted $L_{low}$, $L_{mid}$, and $L_{hi}$ for the lowest, middle, and highest luminosity bins, respectively. We divide each of these nine bins into thirds of $\alpha_{\lambda}$, denoted $\alpha_{low}$, $\alpha_{mid}$, and $\alpha_{hi}$ for the bluest, intermediate, and reddest spectral indices. These cuts are indicated in Figure~\ref{fig:histmolds}.
We do not require the sample to be symmetric in both dimensions, so the ranges for $\alpha_{\lambda}$ in the luminosity bins vary.
The lowest luminosity bin's $\alpha_{\lambda}$ ranges are shifted by $-0.04$ 
from the middle luminosity bin's $\alpha_{\lambda}$ ranges,
and the highest luminosity bin's $\alpha_{\lambda}$ ranges are shifted by $0.04$. 
This asymmetry between bins is exacerbated when considering the tails of the parameter's distributions
and is particularly prevalent in spectral index. 
Thus when we explore the variations in composite spectra along one dimension,
we utilize the middle bins ($\rm mid$) of the fixed parameter because they are more consistent throughout.
By design, the variability across redshift bins is much smaller and neglected.
{ The median values of redshift, $L_{bol}$, and $\alpha_{\lambda}$ that comprise each of these bins is summarized in Table~\ref{tab:bintable}.  In addition, we provide the number of spectra, median FWHM of the \textsc{C iv} emission line, and median signal-to-noise ratio per pixel in the region used to determine bolometric luminosity.  The composite spectra representing either the middle bin in luminosity or the middle bin in spectral index are each derived from more than 1,000 individual spectra.  The signal-to-noise ratio only varies by a factor of 1.5 between the lowest redshift bin and the highest redshift bin for any given luminosity bin.}

We generate composite spectra from the quasars in each of the 27 bins. Spectra were first de-redshifted and resampled to a
common wavelength grid. These spectra were normalized by the median flux in the same wavelength range used to find
$L_{bol}$ ($1340 \le \lambda \le 1360$ \AA). Pixels with corrupted flux (such as those from poorly subtracted sky features)
are masked as described in \citet{harris16a}. At every wavelength pixel, we simply take the median from all quasars
that contribute unmasked flux after removing outliers more than three standard deviations from the median.
Statistical errors for the measured flux are estimated through bootstrap methods.
The resulting high signal-to-noise ratio composite spectrum in each bin represents the specific subset of the
quasar population from which we study spectral variation and evolution.

\begin{figure*}
\centering
\includegraphics[width=\textwidth,height=.52\textwidth,trim=1cm 12.8cm 1cm 3.8cm,clip]{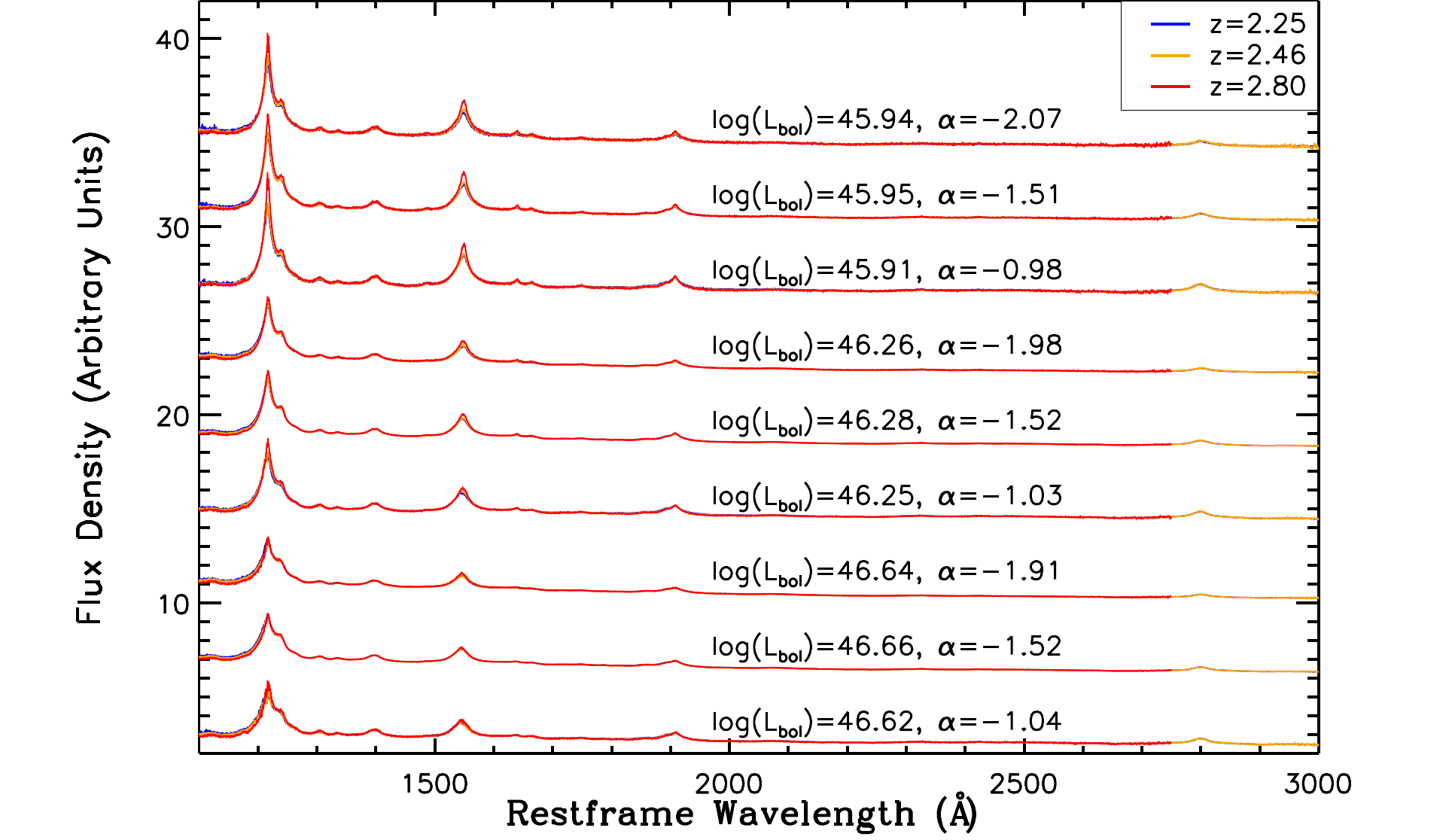}
\caption{\label{fig:comp27} The 27 composite spectra derived from 58,656 objects binned with respect to $\alpha_{\lambda}$ and $L_{bol}$. Each composite is offset by four arbitrary units to avoid confusion. The values reported next to each set of three composite spectra represent the mean luminosity and spectral index over the full redshift range.}
\vspace{-.7cm}
\includegraphics[width=\textwidth,height=.52\textwidth,trim=1cm 12.8cm 1cm 2.6cm,clip]{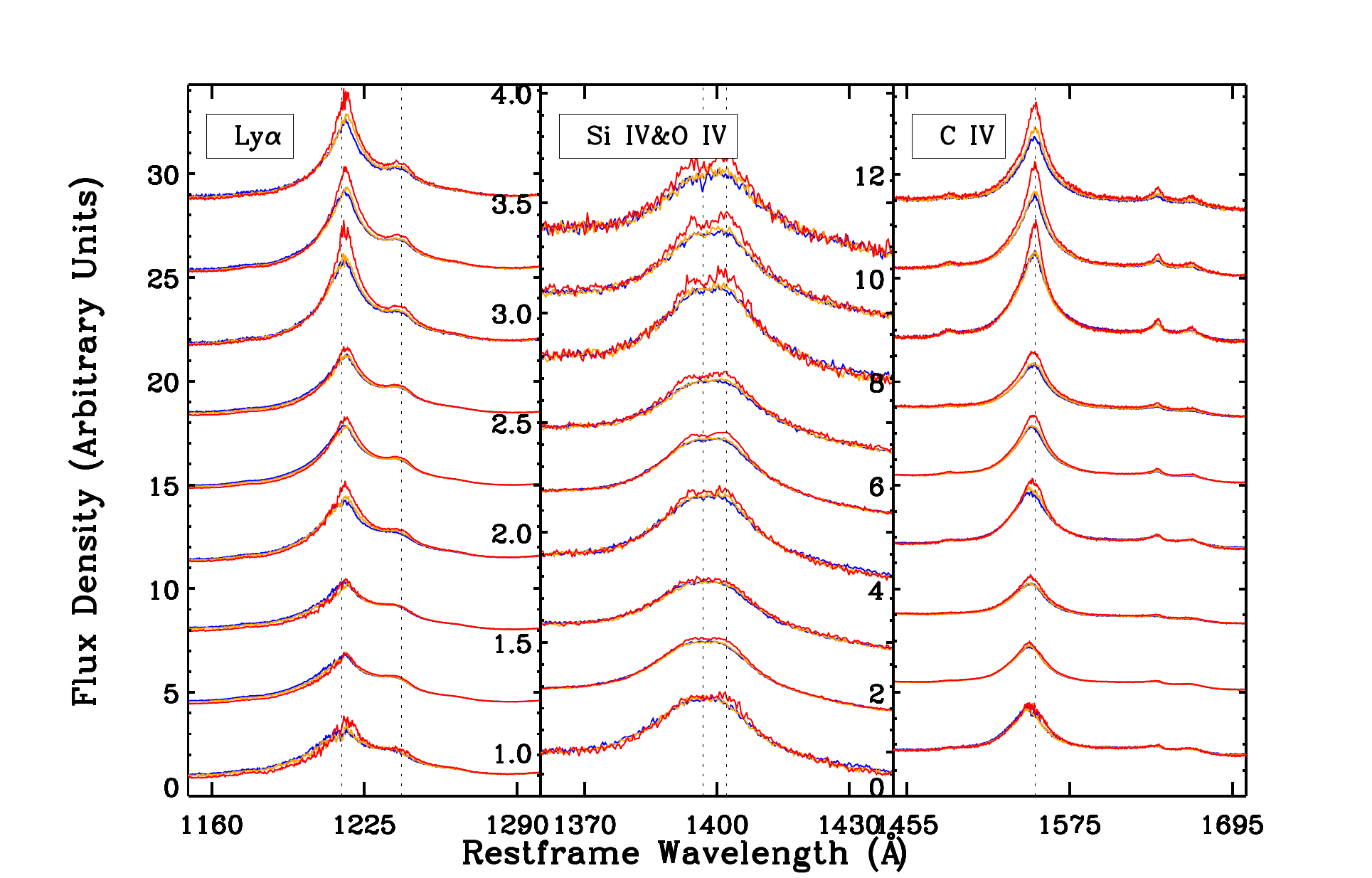}
\caption{\label{fig:comp27zoom} The expanded view of the 27 composite spectra centered on prominent spectral features. The order and color scheme is the same as in Figure~\ref{fig:comp27}. The dashed lines reflect positions of the central wavelengths of emission lines; from left to right they are Ly$\alpha$ $\lambda$1216, \textsc{N v} $\lambda$1240, Si \textsc{iv} $\lambda$1397, \textsc{O iv} $\lambda$1402, and \textsc{C iv} $\lambda$1549. Each composite is offset by $3.5$, $0.3$, and $1.25$ from left to right, respectively.}

\end{figure*}

The composite spectra, ordered from low luminosity at the top to high luminosity at the bottom, are displayed in Figure~\ref{fig:comp27}. The traditional BE is evident in the presence of strong emission features at the top of the figure and the weaker emission features at the bottom. Figure~\ref{fig:comp27zoom} represents an expanded view of the composite spectra in three wavelength regions featuring interesting emission features. There is clear redshift evolution in the line strengths even when fixing luminosity and spectral index.

\clearpage

 


\section{\textbf{Variation in Spectral Signature}}\label{sec:baldwin}
The 27 composite spectra are used to compare changes in spectral features associated with distinctly different quasar populations. Inspecting the traditional parameter used in BE analyses, we investigate equivalent widths ($W_{\lambda}$) for lines of interest in each of the composite spectra. We calculate $W_{\lambda}$ for \textsc{C iv} because it was most commonly used in previous analyses. In addition, we calculate  $W_{\lambda}$ for the Si \textsc{iv} \& \textsc{O iv} blend located at approximately 1400 \AA; this feature is relatively isolated from neighboring lines, and, at 33.5 and 113.9 eV, the ionization potential is significantly different than the 47.9 eV potential of \textsc{C iv} (see Table~\ref{tab:lines}). We could have chosen other lines, such as \textsc{C iii}; however, the other strong lines tend to have complex blends where it is difficult to fit the continuum.

\begin{table}[b]
\caption{\label{tab:lines} Common emission lines, their wavelengths, and the ionization energy required to create each species of ion.}
\begin{center}
\begin{tabular}{l c c}
\hline\hline
Line & Wavelength (\AA) & Ionization Energy (eV)  \\
\hline
O IV & 1402 & 113.9 \\
N V & 1240 & 77.7 \\
C IV & 1549 & 47.9 \\
N IV] & 1486 & 47.4 \\
O III] & 1663 & 35.1 \\
Si IV & 1397 & 33.5 \\
Al III & 1857 & 28.4 \\
C III] & 1909 & 24.4 \\
Si III] & 1892 & 16.3 \\
Ly$\alpha$ & 1216 & 13.6 \\
Mg II & 2798 & 7.6 \\
\hline
\end{tabular}
\end{center}
\end{table}

When computing the equivalent width, we first fit a powerlaw to the estimated continuum in the vicinity of the relevant lines. For \textsc{C iv}, we use the interval [1445,1465] \AA\ and [1700,1705] \AA\ to determine the continuum; for Si \textsc{iv} \& \textsc{O iv}, we use the interval [1360,1370] \AA\ and [1425,1440] \AA. We determine the line flux via a discrete integral computed after continuum subtraction. For \textsc{C iv} we integrate the relative flux density over the region 1500 to 1600 \AA, and for  Si \textsc{iv} \& \textsc{O iv} we use 1370 to 1425 \AA.

\begin{figure*}
\centering
\includegraphics[trim=1.25cm 12.5cm 1cm 10cm,clip]{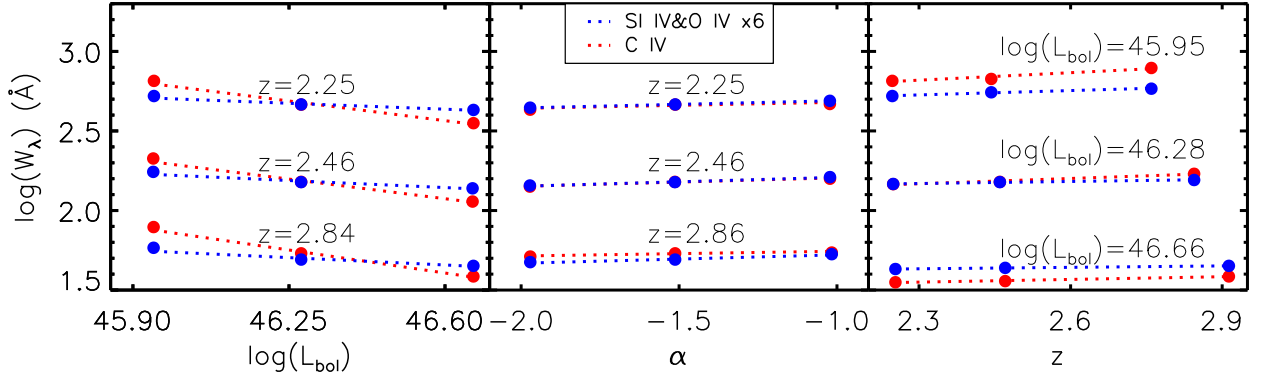}
\caption{The equivalent width as a function of $\textrm{log}(L_{bol})$ (left panel), $\alpha_{\lambda}$ (middle panel), and $z$ (right panel) for the composite spectra. The filled red circles correspond to the equivalent widths measured for \textsc{C iv} and the dotted red line corresponds to the best linear fit. The blue color scheme corresponds to Si \textsc{iv} \& \textsc{O iv}. The Si \textsc{iv} \& \textsc{O iv} equivalent widths are much weaker and are scaled by a factor of six for illustrative purposes. Each series is offset by 0.5 in $\textrm{log}(W_{\lambda})$ for clarity. Error bars are included for all measurements, but are smaller than the symbols that represent the data.}
\label{fig:sepew}
\end{figure*}

Figure~\ref{fig:sepew} presents $\textrm{log}(W_{\lambda})$ as a function of $\textrm{log}(L_{bol})$, $\alpha_{\lambda}$, and $z$. Because of the remaining diversity in the tails of the spectral index distribution described at the end of Section~\ref{subsec:binning}, only the results for $L_{bol}$ and redshift using the intermediate bin in spectral index ($\alpha_{\rm mid}$) are shown. When displaying the results of  $W_{\lambda}$ versus $\alpha_{\lambda}$, we use the intermediate bin in luminosity ($L_{\rm mid}$). Thus the equivalent widths of only nine of the 27 composite spectra are presented in each panel.

Following \citet{vbaldwin77}, we fit the functional form $W_{\lambda} \propto L^{\beta_{w}}$ to the observed $\textrm{log}(W_{\lambda})$/$\textrm{log}(L_{bol})$ relationship. The results of the fit for each redshift bin are shown in Figure~\ref{fig:sepew} and found in Table~\ref{tab:ewfits}. The slope of the relationship is roughly a factor of three lower for the Si \textsc{iv} \& \textsc{O iv} blend than it is for \textsc{C iv}. This result is expected if Si \textsc{iv} is the dominant line in the complex, as the higher ionization potential species display a greater anti-correlation with luminosity \citep[see review by][]{shields07a}. The \textsc{C iv} slopes here are steeper than the results from the application of the BCES$(Y|X)$ method used in \cite{bian12a},
 where they find $\beta_w = -0.238 \pm 0.040$. The slope of equivalent width with luminosity grows steeper by $\sim 20\%$ as the redshift increases from $z = 2.25$ to $z = 2.84$. The same trend is seen in \textsc{C iv} and Si \textsc{iv} \& \textsc{O iv}. \cite{bian12a} report hints of an opposite trend with redshift; however, they acknowledge that the dynamic range in luminosity is quite different across the redshift range of their sample, thus possibly affecting the linear regression. Because we have carefully controlled the sample to have the same range of luminosity in each redshift bin, our analysis should be less susceptible to such a bias.

We next explore the $W_{\lambda}$ correlation with $\alpha_{\lambda}$. As seen in the second panel of Figure~\ref{fig:sepew}, the slope of the relation $\textrm{log}(W_{\lambda}) \propto \alpha_{\lambda}m_{\alpha}$ is much shallower for spectral index than it is for luminosity, even though the range of both parameters spans nearly the entire population of quasars in our sample. From Table~\ref{tab:ewfits}, it is also evident that the equivalent width of Si \textsc{iv} \& \textsc{O iv} has a stronger dependence on spectral index than does the equivalent width of \textsc{C iv}.

Surprisingly, in the third panel of Figure~\ref{fig:sepew}, there is redshift evolution in equivalent width at fixed luminosity and spectral index. The equivalent widths appear correlated with redshift following the relation $\textrm{log}(W_{\lambda}) \propto zm_{z}$, whereby equivalent width is increasing with increasing redshift. The slope of the fit for each luminosity bin is listed in Table~\ref{tab:ewfits}. As with the equivalent width-luminosity relationship, the trend is much steeper for \textsc{C iv} than it is for Si \textsc{iv} \& \textsc{O iv}. Also evident in the table, the trend steepens with decreasing luminosity. Preliminary findings of a similar nature were reported in \cite{vandenberk04a}, but no conclusive findings have been subsequently reported.

We expand upon the correlations with equivalent width by examining spectral diversity and evolution as a function of wavelength in the high signal-to-noise ratio composite spectra. We explore diversity through differential spectra used to provide a pixel-by-pixel linear interpolation between composite spectra. The differential spectra are computed $\frac{\Delta f}{\Delta x} = \frac{f(\lambda)-f(\lambda)_{mid}}{x-x_{mid}}$ for each observed parameter $x = \textrm{log}(L_{bol})$, $\alpha_{\lambda}$, and $z$. As above, we determine $\frac{\Delta f}{\Delta \textrm{log}(L_{bol})}$ and $\frac{\Delta f}{\Delta z}$ only for the composite spectra constructed with the intermediate spectral index ($\alpha_{mid}\approx -1.52$). For $\frac{\Delta f}{\Delta \alpha}$, we use the intermediate luminosity bin ($L_{mid}\approx 46.26$). We discuss the results for each parameter in the following subsections.

\begin{table*}
\centering
\caption{\label{tab:ewfits} Best fit linear slopes from Figure \ref{fig:sepew}}
\begin{tabular}{| c | c c c c c c c c |}
\hline\hline
Feature & Mean $z$ & $\beta_{w}$\tablenotemark{a}  &      & Mean $z$ & $m_{\alpha}$\tablenotemark{b} &      & Mean $\textrm{log}(L_{bol})$ & $m_{z}$\tablenotemark{c} \\
\hline
(\textsc{C iv}) & 2.25 & -0.35 $\pm$ 0.004 & \empty\empty\empty & 2.25 & 0.038 $\pm$ 0.004 & \empty\empty\empty & 45.95 & 0.16 $\pm$ 0.008 \\
(\textsc{C iv}) & 2.46 & -0.35 $\pm$ 0.005 & \empty & 2.46 & 0.053 $\pm$ 0.005 & \empty & 46.28 & 0.11 $\pm$ 0.004 \\
(\textsc{C iv}) & 2.84 & -0.41 $\pm$ 0.005 & \empty & 2.86 & 0.030 $\pm$ 0.006 & \empty & 46.66 & 0.06 $\pm$ 0.004 \\
\hline
(Si \textsc{iv} \& \textsc{O iv}) & 2.25 & -0.11 $\pm$ 0.007 & \empty & 2.25 & 0.045 $\pm$ 0.008 & \empty & 45.95 & 0.091 $\pm$ 0.020 \\
(Si \textsc{iv} \& \textsc{O iv}) & 2.46 & -0.13 $\pm$ 0.008  & \empty & 2.46 & 0.056 $\pm$ 0.008 & \empty & 46.28 & 0.043 $\pm$ 0.007\\
(Si \textsc{iv} \& \textsc{O iv}) & 2.84 & -0.13 $\pm$ 0.009 & \empty & 2.86 & 0.053 $\pm$ 0.010 & \empty & 46.66 & 0.030 $\pm$ 0.006 \\
\hline
\end{tabular}
\tablenotetext{a}{Defined by $W_\lambda \propto L^{\beta_w}$}
\tablenotetext{b}{Defined by $\textrm{log}(W_{\lambda})  \propto \alpha_{\lambda}m_{\alpha}$}
\tablenotetext{c}{Defined by $\textrm{log}(W_{\lambda}) \propto zm_{z}$}
\end{table*}

\FloatBarrier
\subsection{Relationship Between Luminosity and Line Features}\label{subsec:lbol}
\begin{figure*} 
\centering
\includegraphics[width=\textwidth,trim=1.6cm 12.3cm 2.5cm 2.5cm,clip]{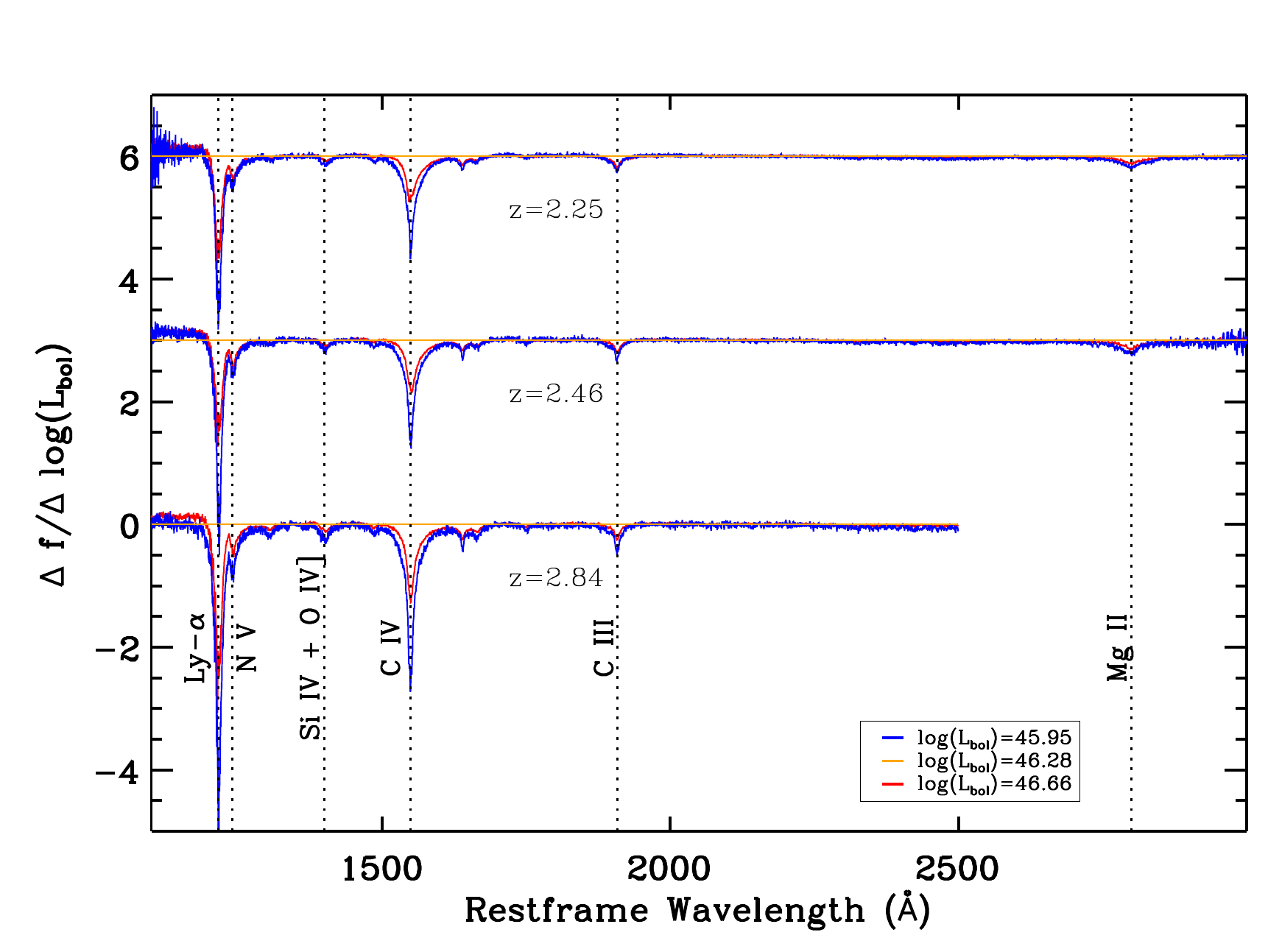}
\caption{
Nine $\frac{\Delta f}{\Delta \textrm{log}(L_{bol})}$ differential spectra derived from composite spectra at fixed spectral index and redshift. Emission lines of interest are labeled and marked with dotted lines. The $\alpha_{\rm mid}$ bin is used for the spectral index with a mean of $\alpha_{\lambda} = -1.52$. To reduce confusion, the differential spectra of each redshift interval are vertically offset by $0.5$ units.}
\label{fig:complbol}
\end{figure*}

The differential spectra representing $\frac{\Delta f}{\Delta \textrm{log}(L_{bol})}$ are shown in Figure~\ref{fig:complbol}. Here, the traditional BE is presented as an inverse correlation between $f(\lambda)$ and luminosity, i.e., $\frac{\Delta f}{\Delta \textrm{log}(L_{bol})}$ is negative in the vicinity of all strong emission lines. The biggest effect is seen in Ly-$\alpha$ and \textsc{C iv}. Smaller effects exist around the \textsc{C iii} $\lambda$1908 blend and Mg \textsc{ii} $\lambda$2799. Although Ly-$\alpha$ displays the strongest effect, most likely due to its intrinsic strength, the smaller effects in \textsc{C iii} and Mg \textsc{ii} are consistent with the trend that lower ionization potential species display a smaller anti-correlation.  \textsc{N v} is a possible exception to this general trend \citep[as outlined in][]{shields07a}, but the linear variation shown in $\frac{\Delta f}{\Delta \textrm{log}(L_{bol})}$ appears fairly strong relative to the intrinsic line strength.

When examining $\frac{\Delta f}{\Delta \textrm{log}(L_{bol})}$ over the three redshift bins, one clearly sees more pronounced emission line features at higher redshift. In the differential spectra representing the linear change from $\textrm{log}(L_{bol})=45.95$ to 46.28, the peak in \textsc{C iv} increases by $\sim 60\%$ from $z=2.25$ to $z=2.84$. Similar changes in structure are present in the other prominent lines.

There is an asymmetry in the wings of \textsc{C iv} in Figure~\ref{fig:complbol}; the red side of the line shows more integrated flux than the blue side. This asymmetry leads to a systematic suppression of flux on the red side of the emission feature with increasing luminosity.
The resulting blueshift of the central wavelength of \textsc{C iv} as a function of luminosity is clear in the
right panel of Figure~\ref{fig:comp27zoom}. \cite{richards11} report a related trend that higher \textsc{C iv} blueshifts are correlated
with higher luminosity sources. 

Finally, there is an asymmetry in the differential spectra determined between $\textrm{log}(L_{bol})=45.95$ to 46.28 and $\textrm{log}(L_{bol})=46.28$ to 46.66. There is stronger variation in lines when interpolating over the lower luminosity range than over the higher luminosity range in each redshift bin.

The signature of $\frac{\Delta f}{\Delta \textrm{log}(L_{bol})}$ bears striking resemblance to the first eigenspectra derived from principal component analysis (PCA) in the results of \cite{shang04a}, \cite{suzuki06a}, and \cite{paris11a}. \cite{shang04a} report that their first eigenspectrum is correlated with luminosity in a similar fashion to the BE. In all of these works, the first eigenspectrum possesses the same sign of variation across all emission lines, as displayed in Figure~\ref{fig:complbol}.


\FloatBarrier
\subsection{Signature of Spectral Index}\label{subsec:alpha}
\begin{figure*} 
\centering
\includegraphics[width=\textwidth,trim=1.6cm 12.3cm 2.5cm 2.5cm,clip]{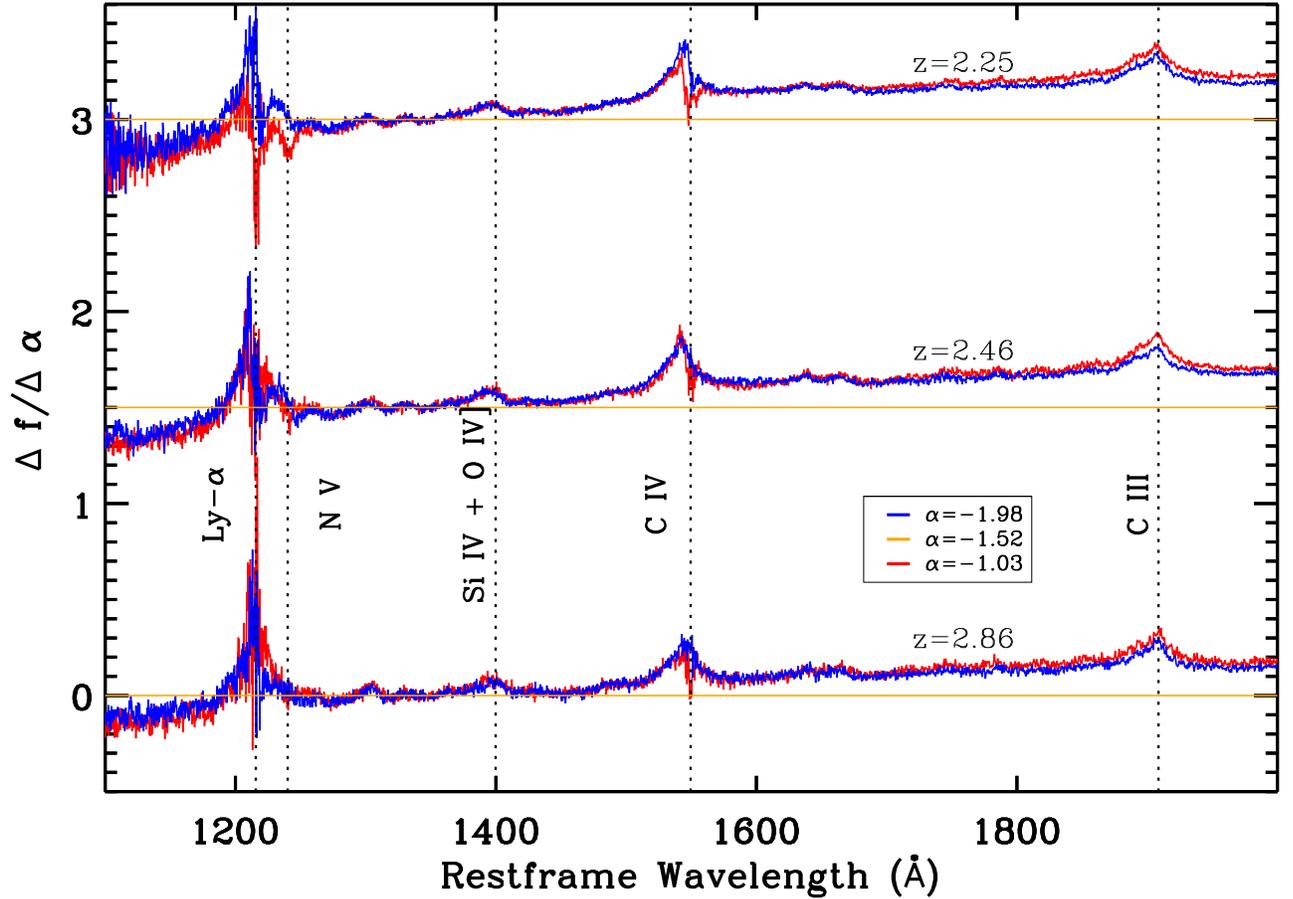}
\caption{Nine $\frac{\Delta f}{\Delta \alpha}$ differential spectra derived from composite spectra at fixed luminosity and redshift. Emission lines of interest are labeled and marked with dotted lines. The $L_{\rm mid}$ bin is used for the luminosity with a mean of $\textrm{log}(L_{bol}) = 46.26$. To reduce confusion, the differential spectra of each redshift interval are vertically offset by $2$ units.}
\label{fig:compalpha}
\end{figure*}

At the beginning of Section~\ref{sec:baldwin}, we found a trend of equivalent width with spectral index,
where the red quasars show stronger emission relative to continuum than the blue quasars.
The analysis of the differential spectra presented in Figure~\ref{fig:compalpha} reveals the same trend that
all the emission lines vary in flux in a correlated fashion, increasing in flux with redder spectra ($\Delta \alpha > 0$).
Deeper investigation shows that the lines in the differential spectra are shifted blueward of the central wavelength.
For example, we measure an average $\sim 1700$  $\textrm{km s}^{-1}$ outflow within \textsc{C iv} per unit change in $\alpha_{\lambda}$.
However, the center of the lines displayed in Figure~\ref{fig:comp27zoom}
do not suggest any noticeable change in the peak wavelength associated with changing spectral index.
The trend seen in Figure~\ref{fig:compalpha} therefore indicates that the strength of a second
outflowing component associated with these emission lines may vary with spectral index.
Also evident is a stronger blueshifting with higher ionization potential species. \textsc{N v} is blueshifted on average by $\sim 2400$ $\textrm{km s}^{-1}$ per unit change in spectral index,
and Ly-$\alpha$ only by $\sim 1400$ $\textrm{km s}^{-1}$,
although estimating the latter offset may be more difficult due to the Ly~$\alpha$ forest.
\textsc{O iv} is heavily blended with Si \textsc{iv}, making it even harder to decouple the two.

A caveat in the analysis of $\frac{\Delta f}{\Delta \alpha}$ is the appearance of some discrepancies between redshift bins. For example, $\frac{\Delta f}{\Delta \alpha}$ computed over the reddest composite spectra at $z=2.25$ shows an absorption profile in \textsc{C iv} that does not appear in any other $\frac{\Delta f}{\Delta \alpha}$ interval. Likewise, the signature of $\frac{\Delta f}{\Delta \alpha}$ varies considerably around \textsc{N v}. These levels of discrepancy do not appear in the modes of variation on luminosity. It is possible that the spectral index analysis is susceptible to systematic errors in the redshift estimation \citep[see][]{denney16a}, calibration, or modeling the continuum. We briefly explored the signature of $\frac{\Delta f}{\Delta \alpha}$ with other techniques of removing or distorting the continuum and found similar discrepancies. For now, we highlight the interesting features in $\frac{\Delta f}{\Delta \alpha}$, but refrain from deeper analysis until a later date.

 \FloatBarrier
\subsection{Redshift Evolution}\label{subsec:redshiftev}
\begin{figure*} 
\centering
\includegraphics[width=\textwidth,trim=1.6cm 12.3cm 2.5cm 2.5cm,clip]{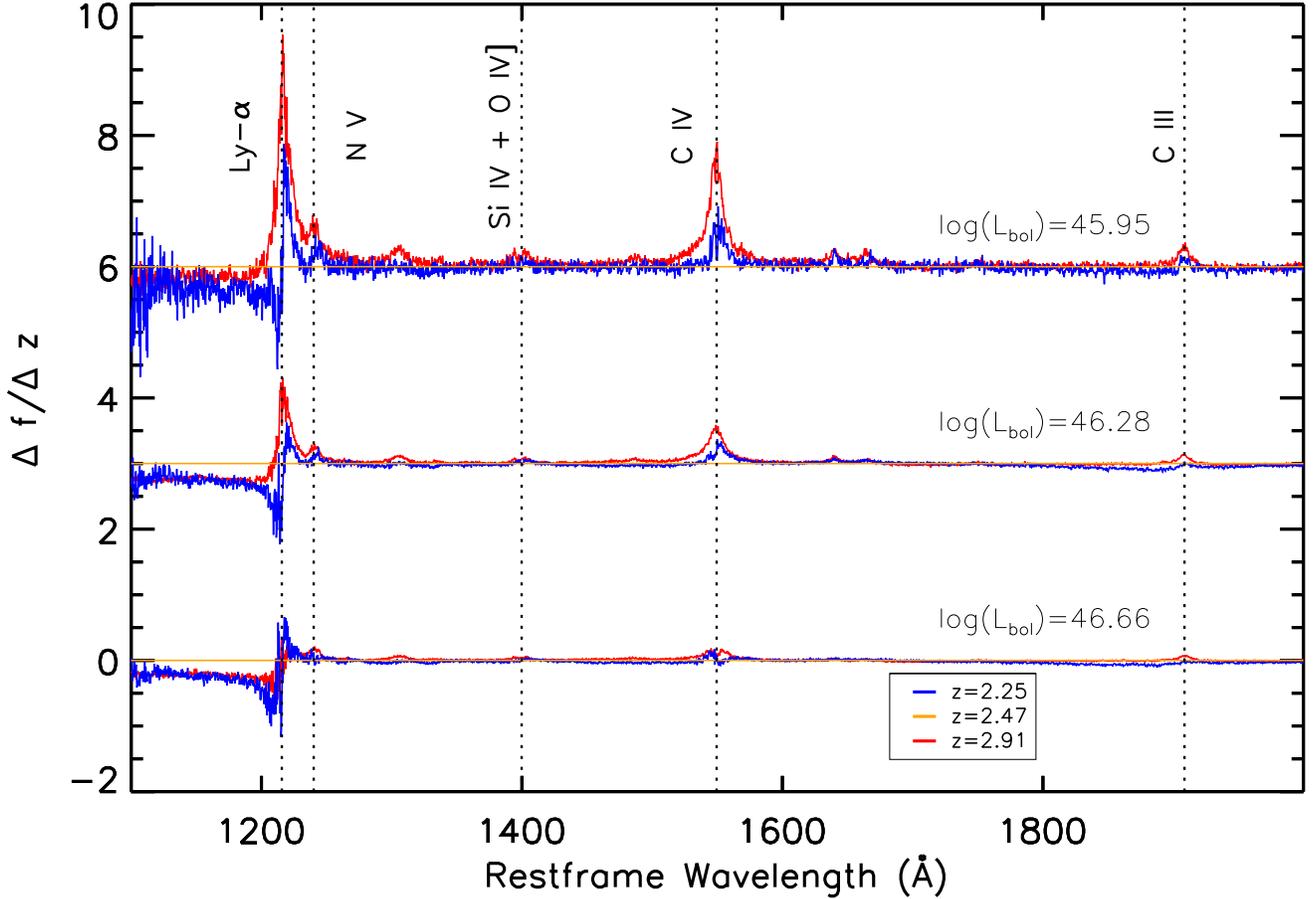}
\caption{Nine $\frac{\Delta f}{\Delta z}$ differential spectra derived from composite spectra at fixed spectral index and luminosity. Emission lines of interest are labeled and marked with dotted lines. The $\alpha_{\rm mid}$ bin is used for the spectral index with a mean of $\alpha_{\lambda} = -1.52$. To reduce confusion, the differential spectra of each redshift interval are offset by $3$ units.}
\label{fig:compz}
\end{figure*}

As shown in Section~\ref{sec:baldwin} and Table~\ref{tab:ewfits}, the equivalent width of both \textsc{C iv} and the Si \textsc{iv} \& \textsc{O iv} complex change with redshift in a manner that is quite similar to the change with luminosity. An examination of the differential spectra representing $\frac{\Delta f}{\Delta z}$, shown in Figure~\ref{fig:compz}, displays this trend in more detail. The behavior of equivalent width with redshift appears as a positive correlation between redshift and the surrounding flux density of each emission line. As with $\frac{\Delta f}{\Delta \textrm{log}(L_{bol})}$, the amplitude of the correlation is the strongest around Ly-$\alpha$ and \textsc{C iv}.

Just as $\frac{\Delta f}{\Delta \textrm{log}(L_{bol})}$ displayed a higher amplitude of the correlation at high redshift,
the $\frac{\Delta f}{\Delta z}$ differential spectra show a stronger correlation at low luminosity.
Comparing the highest luminosity bin to the lowest luminosity bin of the higher redshift interpolation,
reveals a $\sim 1000\%$ increase in the peak line strength of \textsc{C iv}.
The peak flux density in the correlation of the Lyman-$\alpha$ line increases by $\sim 1100\%$ over the same change of luminosity.
As shown in Table~\ref{tab:bintable}, the signature of $\frac{\Delta f}{\Delta z}$ appears to be related to the change in \textsc{C iv} FWHM.
For the bin corresponding to the lowest luminosity and intermediate spectral index,
the \textsc{C iv} FWHM is 4162, 4018, and 3482 $\textrm{km s}^{-1}$ at $z=$ 2.25, 2.44, and 2.76, respectively.
At the bin corresponding to the highest luminosity and intermediate spectral index, the FWHM is 4866, 4871, and 4547 $\textrm{km s}^{-1}$ at the same redshifts.
The average FWHM increases by 19.5\% from $z=2.76$ to $z=2.25$ for the low luminosity bin, but only increases
by 7\% in the high luminosity bin.

In another similar trend to $\frac{\Delta f}{\Delta \textrm{log}(L_{bol})}$, the differential spectra of $\frac{\Delta f}{\Delta z}$
demonstrate asymmetry in the linear interpolations.  The amplitude of the correlation is stronger when interpolating
over the higher redshift composite spectra than the lower redshift range.
At both low and high luminosity, the average FWHM barely changes between $z=2.25$ and $z=2.44$,
while there is significant change in the average FWHM between $z=2.44$ and $z=2.76$.
The same trends with FWHM can be seen in the $\frac{\Delta f}{\Delta \textrm{log}(L_{bol})}$ differential spectra;
larger changes in \textsc{C iv} FWHM lead to higher amplitudes in the differential spectra.

\begin{figure*} 
\centering
\includegraphics[width=\textwidth,trim=1.5cm 12.9cm 2.5cm 2.6cm,clip]{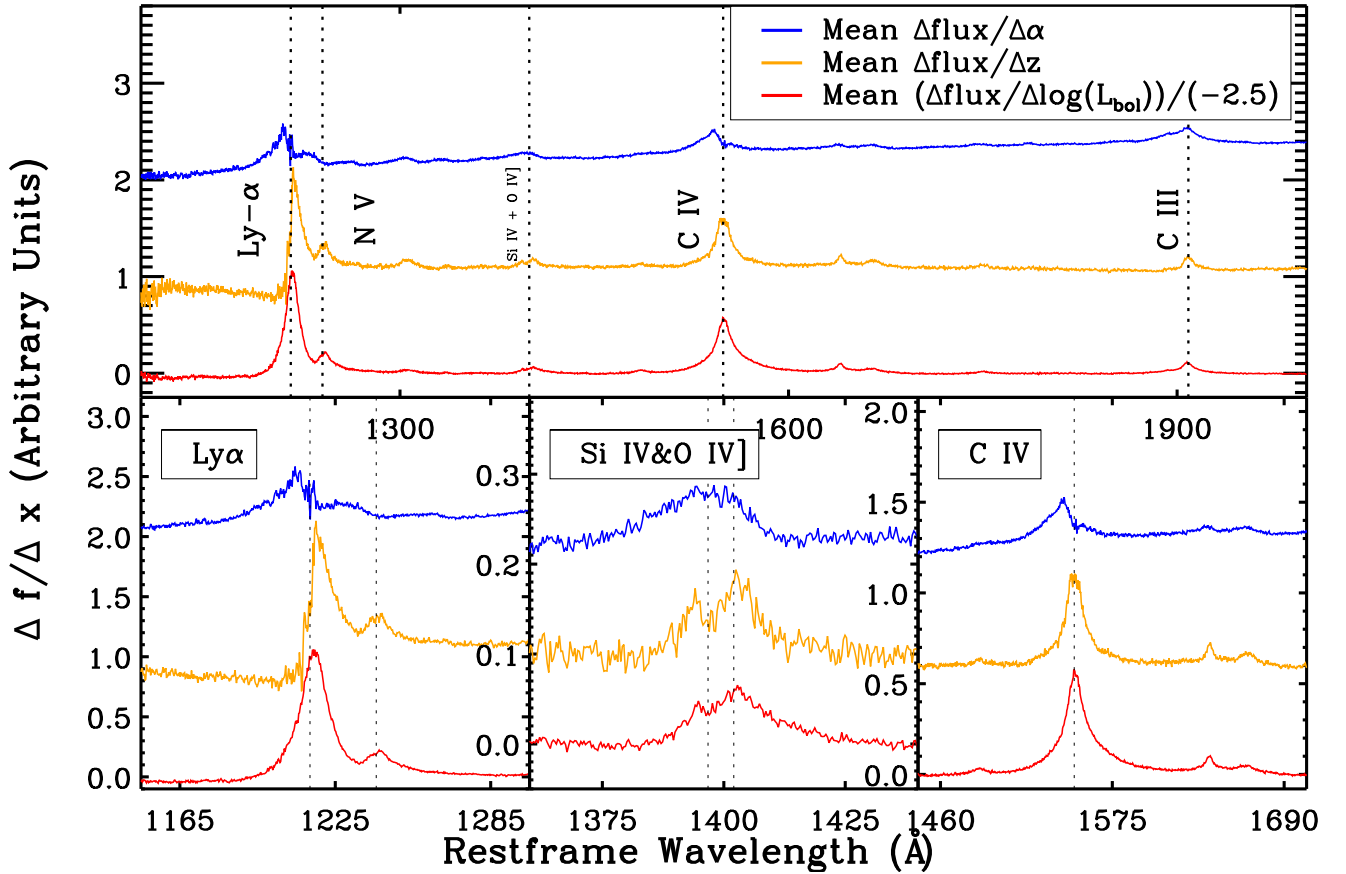}
\caption{Mean of all the differential spectra, where $\frac{\Delta flux}{\Delta \textrm{log}(L_{bol})}$ is divided by $-2.5$ for illustration. The top panel displays differential spectra vertically offset by $1$ arbitrary unit in flux density with emission lines of interest labeled and marked with dotted lines. The bottom set of panels offers extended views of pertinent features with the dotted lines reflecting positions of the central wavelengths of emission lines. From left to right they are Ly-$\alpha$ $\lambda$1216, \textsc{N v} $\lambda$1240, Si \textsc{iv} $\lambda$1397, \textsc{O iv} $\lambda$1402, and \textsc{C iv} $\lambda$1549. Each composite in the bottom panel is offset by $1.0$, $0.1$, and $0.6$ units from left to right, respectively, to reveal comparisons. These figures are designed only to reveal the similarities in the modes, note that the y-axis here is invalid.}
\label{fig:meandiff}
\end{figure*}

The striking resemblance of the redshift signature to the luminosity signature motivates us to quantify the similarity between all differential spectra. To compare the three modes of diversity, we take the mean of all six differential spectra shown in Figure~\ref{fig:complbol} (although inverted to reveal similarities), all six in Figure~\ref{fig:compalpha}, and all six in Figure~\ref{fig:compz}. In doing so, we assume a common mode between all realizations of $\frac{\Delta f}{\Delta x}$ and improve the signal-to-noise ratio for each of the three differential spectra. The high quality differential spectra are shown in Figure~\ref{fig:meandiff}.

There is a strong correlation between $\frac{\Delta f}{\Delta z}$ and $\frac{\Delta f}{\Delta \textrm{log}(L_{bol})}$. The expanded view on Lyman-$\alpha$ shows the same ratio of the change in Lyman-$\alpha$ peak flux relative to \textsc{N v}. The same trend is present when examining the change in the peaks of Si \textsc{iv} and \textsc{O iv} emission. We quantify the correlation between the luminosity and redshift signature by computing the correlation coefficient between the differential spectra. The correlation coefficient is computed over the range $1216$ \AA\ $< \lambda < 2000$ \AA\ to exclude the effects of evolution of the mean optical depth in the Ly-$\alpha$ forest and to avoid introducing noise in the featureless part of the spectra redward of \textsc{C iii}. This analysis produces $\textsc{cor}(\frac{\Delta f}{\Delta z}|\frac{\Delta f}{\Delta \textrm{log}(L_{bol})})= -0.98$, implying that the observed redshift evolution has a nearly identical signature to that of the traditional BE. 

The $\frac{\Delta f}{\Delta \alpha}$ differential spectrum shows the same overall trend that all emission lines vary in the same direction. However, the expanded view of Figure~\ref{fig:meandiff} highlights the blueshifting of emission lines discussed in Section~\ref{subsec:alpha}. The blueshift in $\frac{\Delta f}{\Delta \alpha}$ does not appear in $\frac{\Delta f}{\Delta \textrm{log}(L_{bol})}$, signifying a fundamental difference between these two modes of diversity. Removing the broadband color by distorting the composite spectra to have the same spectral index yields $\textsc{cor}(\frac{\Delta f}{\Delta \alpha}|\frac{\Delta f}{\Delta \textrm{log}(L_{bol})}) = -0.57$.

\FloatBarrier
 
\section{\textbf{Discussion}}\label{discussion}
Having inspected the signatures of luminosity, spectral index, and redshift imprinted on quasar spectra,
we further investigate quasar parameters that may be responsible for the highly correlated
$\frac{\Delta f}{\Delta \textrm{log}(L_{bol})}$ and $\frac{\Delta f}{\Delta z}$ signatures.
The most obvious remaining independent parameters to explore are those associated with emission lines.
Of the strong lines, \textsc{C iv} is both visible at all redshifts covered by this
analysis and not contaminated by absorption in the Ly-$\alpha$ forest.
As shown in Table~\ref{tab:bintable} and discussed in Section~\ref{subsec:redshiftev},
{ the FWHM of the \textsc{C iv} emission line evolves toward smaller values with increasing redshift in each bin of fixed luminosity.}
{ The FWHM in \textsc{C iv} emission decreases by 658 $\textrm{km s}^{-1}$ (16\%) from $z=2.25$ to $z=2.76$ in the lowest luminosity bin. Examination of Figure~\ref{fig:compz} reveals that the highest amplitude of variation with redshift occurs for this low luminosity bin.} Conversely, there is almost no change in the average FWHM over the redshift interval $2.25<z<2.47$ for the highest luminosity bin; the $\frac{\Delta f}{\Delta z}$ computed between these two composite spectra possesses the lowest amplitude of variation. Overall, at a fixed luminosity, the spectroscopic signature of redshift evolution is tightly related to the redshift evolution of \textsc{C iv} FWHM.

Examining the change in \textsc{C iv} FWHM associated with luminosity reveals a similar trend to that in redshift. { At $z=2.25$, the FWHM spans 728 $\textrm{km s}^{-1}$ (17.6\% increase) over the interval $45.95<\textrm{log}(L_{bol})<46.66$. At $z=2.77$, the FWHM increases 1107 $\textrm{km s}^{-1}$ (31.8\%) over the same interval in luminosity. As shown in Table~\ref{tab:ewfits}, the BE slope is significantly steeper at high redshift than it is at low redshift.} Figure~\ref{fig:complbol} shows a clear correlation in the amplitude of $\frac{\Delta f}{\Delta \textrm{log}(L_{bol})}$ with the span in FWHM. { Just as with $\frac{\Delta f}{\Delta z}$, the emission lines in $\frac{\Delta f}{\Delta \textrm{log}(L_{bol})}$ show a larger decrease in strength when there is a larger increase in \textsc{C iv} FWHM.}

\subsection{Spectral Diversity with FWHM}\label{sec:fwhm}
\begin{figure*}
\centering
\includegraphics[width=\textwidth,trim=1.5cm 13cm 2.15cm 3.2cm,clip]{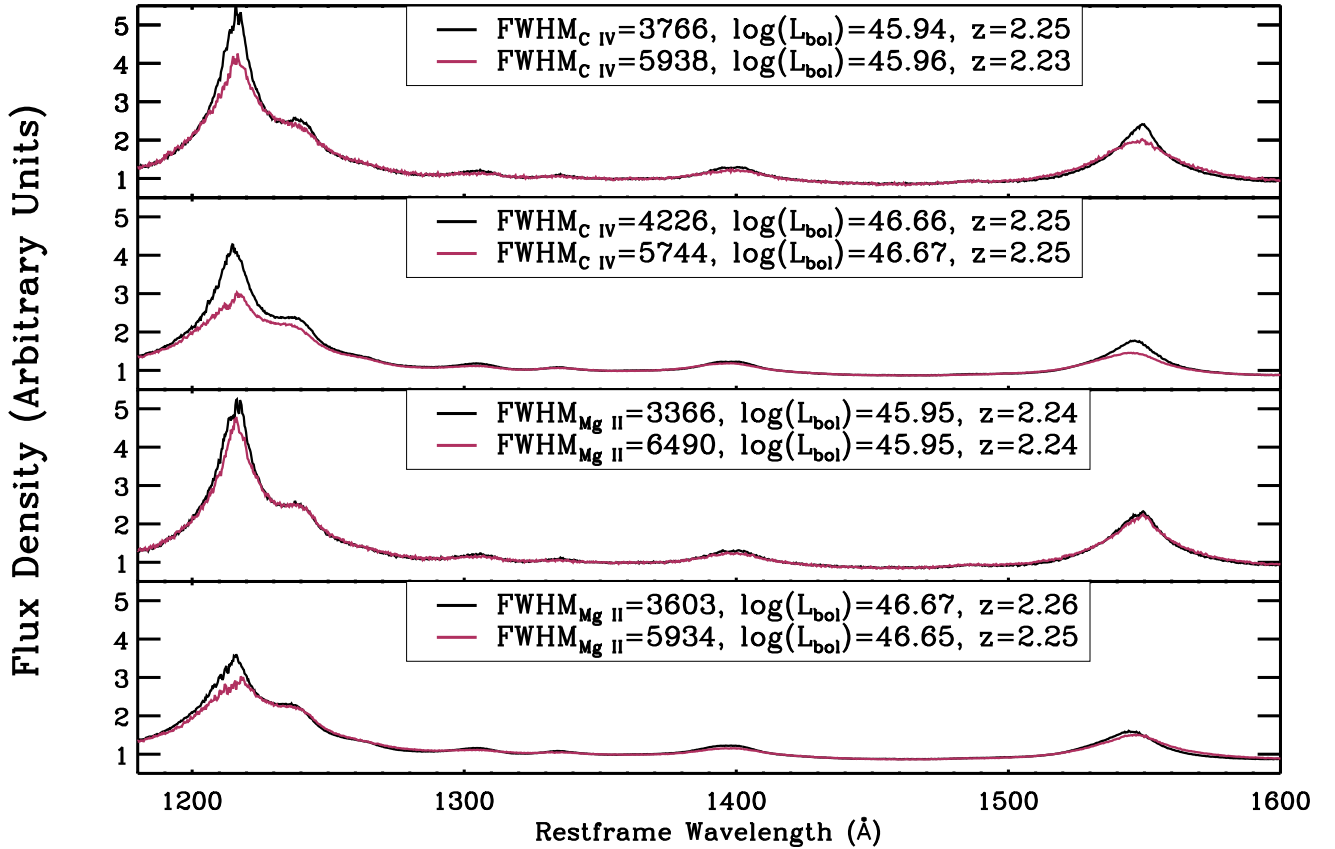}
\caption{The composite spectra binned by FWHM from quasars from the lowest redshift of our sample. Each panel shows the composite spectrum for the quasars with FWHM $<5000 \textrm{km s}^{-1}$ (black) and for quasars with FWHM $>5000 \textrm{km s}^{-1}$ (maroon). The top two panels represent selection according to \textsc{C iv} FWHM and the bottom two panels according to the Mg \textsc{ii} FWHM. Mean FWHM, luminosity and redshift are reported in the legend of each panel. \label{fig:fwhmcomp}}
\end{figure*}

These trends in the amplitude of spectroscopic variations motivate an investigation of the change in spectral profile against FWHM at fixed luminosity. We construct composite spectra by splitting the sample (see Table~\ref{tab:sampletable}) used in Section~\ref{sec:baldwin} by another factor of two. For each $z$, log($L_{bol}$), and $\alpha_{mid}$ bin, we split the sample between quasars with FWHM $< 5000$ $\textrm{km s}^{-1}$ and with FWHM $> 5000$ $\textrm{km s}^{-1}$. To study the dependence on the line chosen for the split, we use both \textsc{C iv} and Mg \textsc{ii} (where available) FWHM. Co-addition of all spectra within each bin results in 18 composite spectra for \textsc{C iv} and six for Mg \textsc{ii}. Examples of the composite spectra in the lowest redshift bin are shown in Figure~\ref{fig:fwhmcomp}.

There is a stronger effect on equivalent width when splitting on \textsc{C iv} than when splitting on Mg \textsc{ii} FWHM. The reasons for this behavior are not obvious and will be the subject of a future work. Focusing on \textsc{C iv} FWHM, there is a correlation of decreased equivalent width with increased FWHM even when holding the luminosity fixed; this is seen strongly in both \textsc{C iv} and Ly-$\alpha$.

\begin{figure*}
\centering
\includegraphics[width=\textwidth,trim=1.5cm 13cm 2.15cm 3.2cm,clip]{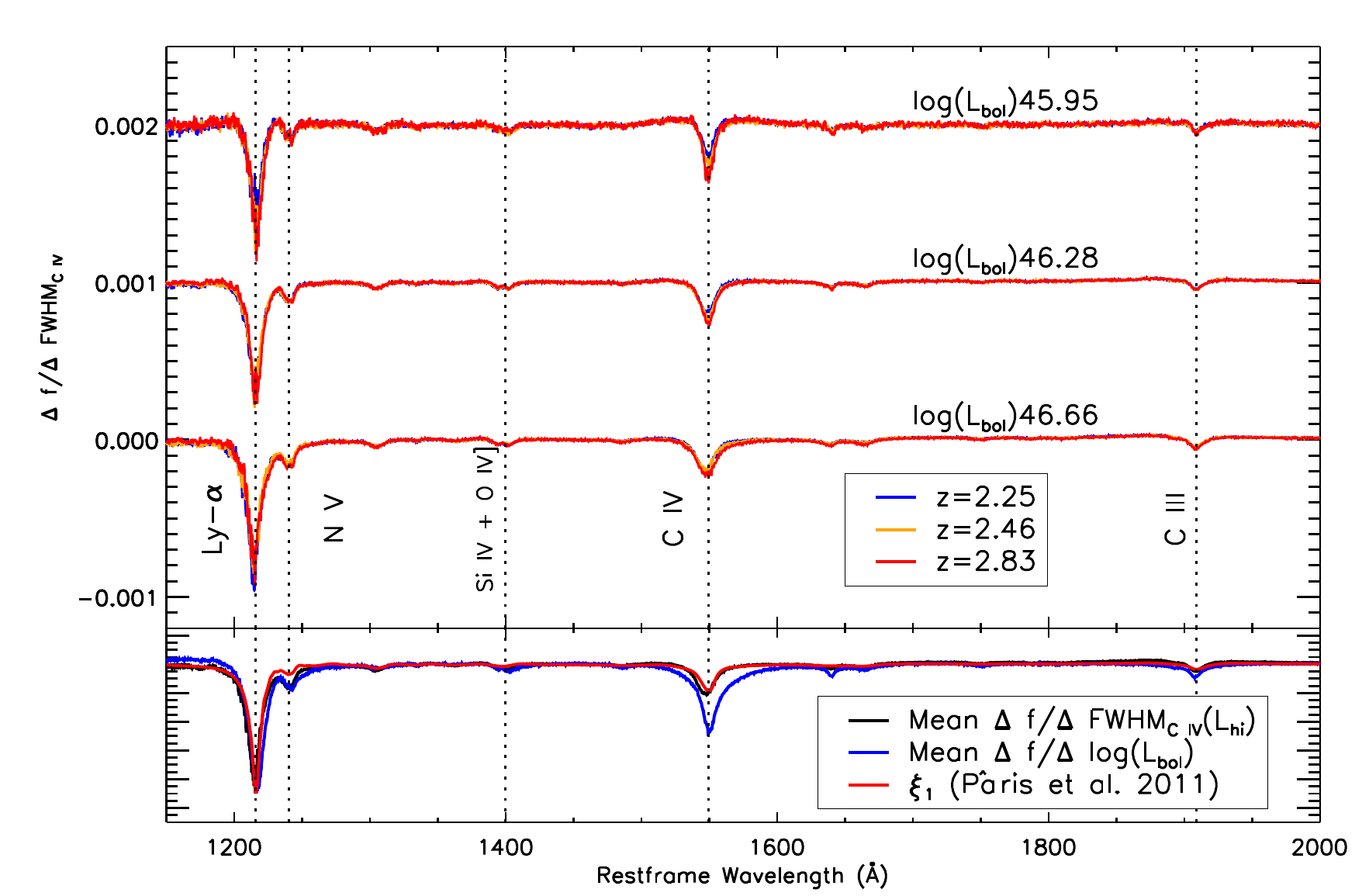}
\caption{\textbf{Top:} Nine $\frac{\Delta f}{\Delta FWHM}$ differential spectra derived from composite spectra at fixed luminosity, spectral index, and redshift. The $\alpha_{\rm mid}$ bin is used for the spectral index with a mean of $\alpha_{\lambda} = -1.52$. The differential spectra computed for each luminosity bin are vertically offset by $0.001$ units for clarity. \textbf{Bottom:} The average $\frac{\Delta f}{\Delta FWHM}$ corresponding to the highest luminosity bin, the average $\frac{\Delta f}{\Delta \textrm{log}(L_{bol})}$, and $\xi_1$ eigenvector from \citet{paris11a}. Emission lines of interest are labeled and marked with dotted lines. A scaling has been applied such that the spectra are normalized to have the same peak in Ly~$\alpha$. \label{fig:fwhmdiff}}
\end{figure*}

For each of the nine bins of fixed luminosity and redshift, we take the differential spectrum representing $\frac{\Delta f}{\Delta FWHM}$ using the \textsc{C iv} FWHM measurements; the results are shown in Figure~\ref{fig:fwhmdiff}. The emission line features are negatively correlated with \textsc{C iv} FWHM, indicating that within bins of fixed luminosity, an increase in \textsc{C iv} FWHM results in a reduction of equivalent width. The differential spectrum for FWHM computed at the highest luminosity bin and the mean differential spectrum for luminosity are compared to the $\xi_1$ eigenvector from \cite{paris11a} in the bottom panel of Figure~\ref{fig:fwhmdiff}. The \cite{paris11a} $\xi_1$ eigenvector more closely matches our $\frac{\Delta f}{\Delta FWHM}$ line ratio between Ly-$\alpha$ and \textsc{C iv} than it matches that line ratio in $\frac{\Delta f}{\Delta \textrm{log}(L_{bol})}$. Computing the correlation coefficients between the $\xi_1$ eigenvector, the FWHM differential spectrum, and the luminosity differential spectrum, the correlations are above 92\% for each of the three pairs. While the correlation coefficients are high for all comparisons, the subtle difference in line ratios (e.g., Ly-$\alpha$/\textsc{C iv}) implies that $\frac{\Delta f}{\Delta \textrm{log}(L_{bol})}$ is a different signature than $\frac{\Delta f}{\Delta FWHM}$, whereas in the $\xi_1$ eigenvector and $\frac{\Delta f}{\Delta FWHM}$ are more closely related.


Of equal importance in Figure~\ref{fig:fwhmdiff}, there is much smaller redshift evolution in the $\frac{\Delta f}{\Delta FWHM}$ differential spectra than in the other differential spectra. For example, the peaks in \textsc{C iv} in $\frac{\Delta f}{\Delta \textrm{log}(L_{bol})}$ change by roughly 60\% between redshift $z=2.25$ and $z=2.84$. In the highest luminosity bin, the peak in \textsc{C iv} in $\frac{\Delta f}{\Delta FWHM}$ only changes by 15\%. The most prominent redshift evolution in the FWHM differential spectra appears at low luminosity, where the peak in \textsc{C iv} changes by 29\% between redshifts $z=2.25$ and $z=2.83$.

{ While the traditional BE is often attributed to a dependence on
luminosity, the spectral variations appear in almost the same fashion
when varying \textsc{C iv} FWHM or redshift while holding luminosity constant.}
The similarity of these trends indicates that the BE is not due to luminosity,
but instead more directly related to \textsc{C iv} FWHM.  The appearance of the BE with luminosity is 
due to the simple fact that that luminosity and FWHM tend to be correlated.
The fact that $\frac{\Delta f}{\Delta FWHM}$ and the $\xi_1$ eigenvector from \cite{paris11a} are so similar
indicates that \textsc{C iv} FWHM is a
larger driver of quasar diversity than any of the parameters we have explored.
We have not navigated the full parameter space of quasar properties, so it is possible that the
$\frac{\Delta f}{\Delta \textrm{log}(L_{bol})}$ signature is a superposition of several effects that include $\frac{\Delta f}{\Delta FWHM}$.

\subsection{{ Physics in the Differential Spectra}}\label{sec:disc2}

The high correlation between the differential spectra associated with luminosity and those with
redshift implies that these variations likely arise from the same physical mechanisms.
In this context, the BE is actually driven by 
some other fundamental quasar parameters that are correlated with luminosity and evolve with redshift.
Black hole mass, Eddington ratio and orientation of the accretion disk are likely physical drivers
for quasar diversity, but are impossible to estimate directly from the high redshift BOSS spectra.
As explained in the previous subsection, the primary driver for spectral diversity in the BOSS quasars is
\textsc{C iv} FWHM rather than luminosity.  
To the extent that \textsc{C iv} FWHM can be used as a proxy for the mass of the black hole,
these correlations can be used to infer the physical mechanisms driving the BE and
redshift evolution in quasar spectra.

The \textsc{C iv} FWHM is known to lead to estimates of black hole mass
with significant scatter in relation to masses derived from
H$\beta$ or H$\alpha$ \citep[e.g.,][]{trakhtenbrot12a, mejia16a}. 
In addition, biases in the FWHM - mass relation arise due to the presence of
outflowing gas which introduces asymmetry and a blueshift with respect to the systemic redshift
\citep[e.g.][]{sulentic07a, netzer07a, brotherton15a,coatman16a}.
As shown in Section~\ref{subsec:alpha}, we have accounted for spectral index which encapsulates some of the outflow signature of \textsc{C iv}.
In addition, we controlled each subsample to have a large number of quasars and identical average luminosity
across all redshift bins.
While it is likely that some biases from outflows associated with \textsc{C iv} FWHM still remain,
the following arguments depend only on the broad assumption that increases (decreases) in \textsc{C iv} FWHM 
at a fixed luminosity represent increased (decreased) black hole mass and decreased (increased) Eddington efficiency.

Keeping the caveats of the exact FWHM - mass relation in mind,
we use the \textsc{C iv} line to evaluate the trends of spectral diversity in terms of changes in black hole mass,
as our redshift range does not provide access to H$\beta$ or \textsc{Mg ii}.
The redshift evolution in average \textsc{C iv} FWHM at a fixed luminosity therefore represents an evolution
toward larger black hole masses at lower redshifts.  
Because luminosity has been factored out of the analysis of redshift evolution, this trend is opposite
to what might be expected for a flux-limited sample.
In the $\frac{\Delta f}{\Delta z}$ differential spectra, 
this evolution appears as a reduction in line strengths relative to continuum at later times.

Using the \textsc{C iv} FWHM measurements as a tracer of the black hole mass, 
the $\frac{\Delta f}{\Delta FWHM}$ differential spectra can be interpreted as an anti-correlation
between black hole mass and \textsc{C iv} equivalent width.
The observed relationship is consistent
with the standard thin accretion disk model \citep{shakura73a}.
The trend we observe can be attributed directly to the 
black hole mass (where the peak temperature of the disk scales as $T_{\rm max} \propto M^{-\frac{1}{4}}$),
or it can be attributed to the Eddington ratio (where the peak temperature of the disk scales as $T_{\rm max} \propto f_{\rm edd}^{\frac{1}{4}}$).
Because we have fixed the luminosity to be constant across the $\frac{\Delta f}{\Delta FWHM}$ and the $\frac{\Delta f}{\Delta z}$ differential spectra,
it is impossible to determine whether the equivalent width is suppressed by a decrease in temperature arising from increased black hole mass
or a decreased Eddington ratio.

Measurements of the intrinsic BE inform the degeneracy in our analysis between
black hole mass and Eddington ratio.
The intrinsic BE can be attributed to
changes in Eddington ratio or other factors that are independent of black hole mass,
however, it is not clear if the relationship stems from the same physical mechanism as the global BE.
As with the global BE, an anti-correlation between \textsc{C iv} equivalent width and continuum luminosity is
seen as the intrinsic BE in multi-epoch spectra of AGN. 
This signature could be attributed to a shielding-gas scenario proposed in \cite{wu11a} \citep[also see][]{luo15a}.
In this model, an increase in the accretion rate (Eddington ratio) is accompanied by an inflation
of the accretion disk which reduces the number of ionizing photons reaching the \textsc{C iv} emitting clouds.
In the scenario that the physics of the intrinsic and global BE are related,
the intrinsic BE predicts a suppression of line strengths with increased Eddington ratio across large quasar samples.
The fact that we observe an opposite trend suggests that the spectroscopic
signature of black hole mass dominates over the spectroscopic signature of Eddington ratio.
Perhaps in the global BE there is a competition between black hole mass and accretion rate,
causing a shallower slope in the luminosity-equivalent width relationship than what is seen in the intrinsic BE.

Following the logic above, the redshift evolution of quasar spectra at fixed luminosity
carries the spectroscopic signature of the global reduction in accretion rate in supermassive black holes.
Assuming a specific relation where the black hole mass at fixed luminosity scales as the square of FWHM,
the average black hole mass in the low luminosity bin increases by about 40\% from $z = 2.77$ to $z = 2.25$,
implying a 30\% global decrease in Eddington ratio at the range of masses probed in this analysis.
The Horizon-AGN simulation \citep[e.g.,][]{dubois14a} suggests that the average 10$^8$ M$_{\odot}$ black hole
decreases by a factor of 4--5 in Eddington ratio between $z = 3$ and $z = 2$ \citep{volunteri16}.
While the simulations cover black holes at all accretion rates rather than the population studied here,
they do indicate that the quasar population is evolving quickly.
This rapid change in Eddington efficiency of the quasar population in the redshift range covered by the
BOSS Ly-$\alpha$ forest quasar sample is apparent in \cite{volunteri16} and provides convincing evidence that there are 
fundamental parameters other than luminosity responsible for quasar diversity.

\FloatBarrier


\section{\textbf{Conclusion }}\label{sec:conclusion}
Using 112,000 quasars covering the redshift range $2.1 \le z \le 3.5$, we sample 27 different bins spanning redshift,
$\textrm{log}(L_{bol})$, and $\alpha_{\lambda}$.   Each of these bins represents an observationally distinct subset of the quasar population.
We shape the distributions of $\textrm{log}(L_{bol})$ and $\alpha_{\lambda}$ to be nearly identical for all redshifts,
resulting in a final sample of 58,656 quasars from which to construct high signal-to-noise ratio composite spectra. 
We derive differential spectra $\frac{\Delta f}{\Delta \textrm{log}(L_{bol})}$, $\frac{\Delta f}{\Delta \alpha}$, 
and $\frac{\Delta f}{\Delta z}$ to describe the linear variations from one composite spectrum to another.
There is a unique signature in $\frac{\Delta f}{\Delta \alpha}$ but we focus on
the signature of diversity associated with luminosity and redshift in this work.  Motivated by redshift evolution in
\textsc{C iv} FWHM, we generate differential spectra $\frac{\Delta f}{\Delta FWHM}$ for direct comparison to the
variations seen with luminosity and redshift. Our primary results are:

\begin{enumerate}
 \item A comparison of the 27 composite spectra confirms the well-documented anti-correlation between equivalent width and luminosity.
Fitting the relation $W_{\lambda} \propto L^{\beta_{w}}$, we find a redshift evolution in the relationship such that
$\beta_w = -0.35 \pm 0.004 , -0.35 \pm 0.005 , \textrm{ and } -0.41 \pm 0.005$ for \textsc{C iv} at $z = 2.25, 2.46, \textrm{ and } 2.84$, respectively.
Higher ionization species typically reflect a larger anti-correlation (e.g., \textsc{C iv} vs Mg \textsc{ii}). 

 \item There is a surprising redshift evolution in quasar spectra at constant luminosity.  The slope of \textsc{C iv} equivalent width as a function of redshift ($\textrm{log}(W_{\lambda}) \propto zm_{z}$) is $m_{z} = 0.16 \pm 0.008 , 0.11 \pm 0.004 , \textrm{ and } 0.06 \pm 0.004$ for $\textrm{log}(L_{bol}) = 45.95, 46.28, \textrm{ and } 46.66$, respectively. 

\item The differential spectrum associated with luminosity correlates with the differential spectrum associated with redshift at 98\%.

\item The average value of \textsc{C iv} FWHM { decreases} significantly with redshift for a given luminosity.
{ The redshift evolution of quasar spectra at a fixed luminosity is directly related to the degree by which \textsc{C iv} FWHM changes between redshift bins.}
The amplitude of spectroscopic variation due to luminosity is also larger for larger spans of \textsc{C iv} FWHM between luminosity bins,
thus leading to a redshift evolution in the slope of the luminosity - equivalent width relationship.

\item The differential spectra that describe the change in flux with respect to \textsc{C iv} FWHM reveal that \textsc{C iv} FWHM
is negatively correlated with equivalent width.
The \textsc{C iv} FWHM differential spectra show a much smaller redshift evolution than do the luminosity differential spectra.

\item The differential spectrum associated with \textsc{C iv} FWHM is not identical to that of the luminosity,
but does correlate at the 93\% level. { The small differences in line features between the two differential spectra imply
that there are other factors beyond FWHM that contribute to the diversity associated with luminosity.}
The \textsc{C iv} FWHM differential spectra bear more resemblance (via the correlation coefficient and comparison of the Ly-$\alpha$/\textsc{C iv} line ratios) to the $\xi_1$ eigenvector derived from high redshift quasar spectra
than to the luminosity differential spectra, implying that the FWHM is largely responsible for the variation in quasar spectra.

\item { Under the basic assumption that \textsc{C iv} FWHM increases with black hole mass
at a fixed luminosity, our analysis shows that the observed BE is not
directly related to luminosity, but instead to a superposition of
spectral diversity associated with black hole mass and Eddington
efficiency. Models for AGN growth predict that average black hole mass
increases and average Eddington efficiency decreases with time.  The
redshift evolution in the BOSS quasar sample at a given luminosity can
therefore be explained by cosmic evolution in the relative contributions
of \textsc{C iv} estimated black hole mass and mass accretion rate.}

 \end{enumerate}
\par

This work presents a perspective on quasar spectra that disentangles several effects responsible for quasar diversity.
The basis chosen here of luminosity, spectral index, and redshift allows identification of clear redshift evolution and demonstrates that luminosity is not the direct cause for the significant variations in line features.
However, the high correlation between the luminosity and redshift differential spectra indicates that the assumed
basis is not optimal for capturing the full diversity in quasar spectra. We have made the 27 median composite spectra and four composite differential spectra available via download at a publicly accessible website\footnote[1]{https://data.sdss.org/sas/dr14/eboss/qso/composite/binned/}.  Future work will focus on the effects of
spectral index and other observable properties (such as FWHM) to isolate the more fundamental drivers of spectral diversity.
The sample chosen here also covers only the redshift range sampled by the BOSS Ly-$\alpha$ forest quasar sample.
The eBOSS program will include an even larger sample of quasars in the $0.9<z<2.2$ redshift range \citep{myers15a},
bringing the total sample to nearly 1,000,000 quasar spectra obtained in the four generations of SDSS.
Future work will include lower redshift quasars and extended rest-frame wavelength coverage that includes the Balmer lines
and the forbidden narrow lines over this much larger sample.
A large number of these quasars also have multi-epoch spectra \citep[e.g.,][]{shen15a}, 
allowing studies of the intrinsic variability of quasars, such as the intrinsic BE.  Such studies may allow a separation of the contributions of mass accretion rate
from the contributions of black hole mass in quasar diversity.
Finally, one of the primary reasons for this analysis is to improve the templates used in redshift
classification for cosmological spectroscopic surveys.  Toward that goal, we will incorporate the 
composite spectra divided by physical properties into the automated redshift pipeline being developed for eBOSS
and calibrate the performance against visual inspections.



\vspace{0.6cm}
\par
Funding for the Sloan Digital Sky Survey IV has been provided by the Alfred P. Sloan Foundation and the Participating Institutions. SDSS-IV acknowledges support and resources from the Center for High-Performance Computing at the University of Utah. The SDSS web site is www.sdss.org.

SDSS-IV is managed by the Astrophysical Research Consortium for the Participating
Institutions of the SDSS Collaboration including the Brazilian Participation Group, Carnegie
Institution for Science, Carnegie Mellon University, the Chilean Participation Group,
Harvard-Smithsonian Center for Astrophysics, Instituto de Astrof\'isica de Canarias,
The Johns Hopkins University, Kavli Institute for the Physics and Mathematics of the
Universe (IPMU) / University of Tokyo, Lawrence Berkeley National Laboratory, Leibniz
Institut f\"ur Astrophysik Potsdam (AIP), Max-Planck-Institut f\"ur Astrophysik (MPA Garching),
Max-Planck-Institut f\"ur Extraterrestrische Physik (MPE), Max-Planck-Institut f\"ur Astronomie
(MPIA Heidelberg), National Astronomical Observatory of China, New Mexico State University,
New York University, University of Notre Dame, Observat\'orio Nacional do Brasil,
The Ohio State University, Pennsylvania State University, Shanghai Astronomical Observatory,
United Kingdom Participation Group, Universidad Nacional Aut\'onoma de M\'exico,
University of Arizona, University of Colorado Boulder, University of Portsmouth,
University of Utah, University of Washington, University of Wisconsin, Vanderbilt University,
and Yale University.

The work of VM, KD, JB, and VK was supported in part by U.S. Department of Energy, Office of
Science, Office of High Energy Physics, under Award Number DE-SC0009959.


\bibliographystyle{mnras} 
\bibliography{bibliography}

\begin{thebibliography}{92}
\expandafter\ifx\csname natexlab\endcsname\relax\def\natexlab#1{#1}\fi

\bibitem[{Alam} et~al.(2015){Alam}, {Albareti}, {Allende Prieto}
  et~al.]{alam15a}
{Alam} S., {Albareti} F.~D., {Allende Prieto} C., et~al., 2015, \apjs, 219, 12

\bibitem[{Alam} et~al.(2016){Alam}, {Ata}, {Bailey} et~al.]{alam16a}
{Alam} S., {Ata} M., {Bailey} S., et~al., 2016, arXiv:1607.03155

\bibitem[{Bachev} et~al.(2004){Bachev}, {Marziani}, {Sulentic}, {Zamanov},
  {Calvani} \& {Dultzin-Hacyan}]{bachev04a}
{Bachev} R., {Marziani} P., {Sulentic} J.~W., {Zamanov} R., {Calvani} M.,
  {Dultzin-Hacyan} D., 2004, \apj, 617, 171

\bibitem[{Baldwin}(1977)]{vbaldwin77}
{Baldwin} J.~A., 1977, \apj, 214, 679

\bibitem[{Baldwin} et~al.(1989){Baldwin}, {Wampler} \& {Gaskell}]{vbaldwin89}
{Baldwin} J.~A., {Wampler} E.~J., {Gaskell} C.~M., 1989, \apj, 338, 630

\bibitem[{Baskin} et~al.(2015){Baskin}, {Laor} \& {Hamann}]{baskin15a}
{Baskin} A., {Laor} A., {Hamann} F., 2015, \mnras, 449, 1593

\bibitem[{Bian} et~al.(2012){Bian}, {Fang}, {Huang} \& {Wang}]{bian12a}
{Bian} W.-H., {Fang} L.-L., {Huang} K.-L., {Wang} J.-M., 2012, \mnras, 427,
  2881

\bibitem[{Bolton} et~al.(2012){Bolton}, {Schlegel}, {Aubourg}
  et~al.]{bolton12a}
{Bolton} A.~S., {Schlegel} D.~J., {Aubourg} {\'E}., et~al., 2012, \aj, 144, 144

\bibitem[{Boroson}(2002)]{boroson02}
{Boroson} T.~A., 2002, \apj, 565, 78

\bibitem[{Boroson} \& {Green}(1992)]{boroson92}
{Boroson} T.~A., {Green} R.~F., 1992, \apjs, 80, 109

\bibitem[{Bovy} et~al.(2011){Bovy}, {Hennawi}, {Hogg} et~al.]{bovy11a}
{Bovy} J., {Hennawi} J.~F., {Hogg} D.~W., et~al., 2011, \apj, 729, 141

\bibitem[{Bovy} et~al.(2012){Bovy}, {Myers}, {Hennawi} et~al.]{bovy12a}
{Bovy} J., {Myers} A.~D., {Hennawi} J.~F., et~al., 2012, \apj, 749, 41

\bibitem[{Brotherton} et~al.(2015){Brotherton}, {Runnoe}, {Shang} \&
  {DiPompeo}]{brotherton15a}
{Brotherton} M.~S., {Runnoe} J.~C., {Shang} Z., {DiPompeo} M.~A., 2015, \mnras,
  451, 1290

\bibitem[{Brotherton} et~al.(2001){Brotherton}, {Tran}, {Becker}, {Gregg},
  {Laurent-Muehleisen} \& {White}]{brotherton01a}
{Brotherton} M.~S., {Tran} H.~D., {Becker} R.~H., {Gregg} M.~D.,
  {Laurent-Muehleisen} S.~A., {White} R.~L., 2001, \apj, 546, 775

\bibitem[{Cackett} \& {Horne}(2006)]{cackett06}
{Cackett} E.~M., {Horne} K., 2006, \mnras, 365, 1180

\bibitem[{Coatman} et~al.(2016){Coatman}, {Hewett}, {Banerji} \&
  {Richards}]{coatman16a}
{Coatman} L., {Hewett} P.~C., {Banerji} M., {Richards} G.~T., 2016, \mnras,
  461, 647

\bibitem[{Cole} et~al.(2005){Cole}, {Percival}, {Peacock} et~al.]{cole05a}
{Cole} S., {Percival} W.~J., {Peacock} J.~A., et~al., 2005, \mnras, 362, 505

\bibitem[{Croom} et~al.(2002){Croom}, {Rhook}, {Corbett} et~al.]{croom02a}
{Croom} S.~M., {Rhook} K., {Corbett} E.~A., et~al., 2002, \mnras, 337, 275

\bibitem[{Dawson} et~al.(2016){Dawson}, {Kneib}, {Percival} et~al.]{dawson16a}
{Dawson} K.~S., {Kneib} J.-P., {Percival} W.~J., et~al., 2016, \aj, 151, 44

\bibitem[{Dawson} et~al.(2013){Dawson}, {Schlegel}, {Ahn} et~al.]{dawson13a}
{Dawson} K.~S., {Schlegel} D.~J., {Ahn} C.~P., et~al., 2013, \aj, 145, 10

\bibitem[{Delubac} et~al.(2015){Delubac}, {Bautista}, {Busca}
  et~al.]{delubac15a}
{Delubac} T., {Bautista} J.~E., {Busca} N.~G., et~al., 2015, \aap, 574, A59

\bibitem[{Denney} et~al.(2016){Denney}, {Horne}, {Brandt} et~al.]{denney16a}
{Denney} K.~D., {Horne} K., {Brandt} W.~N., et~al., 2016, arXiv:1605.08057

\bibitem[{Dietrich} et~al.(2002){Dietrich}, {Hamann}, {Shields}
  et~al.]{dietrich02a}
{Dietrich} M., {Hamann} F., {Shields} J.~C., et~al., 2002, \apj, 581, 912

\bibitem[{Dubois} et~al.(2014){Dubois}, {Volonteri} \& {Silk}]{dubois14a}
{Dubois} Y., {Volonteri} M., {Silk} J., 2014, \mnras, 440, 1590

\bibitem[{Eisenstein} et~al.(2011){Eisenstein}, {Weinberg}, {Agol}
  et~al.]{eisenstein11a}
{Eisenstein} D.~J., {Weinberg} D.~H., {Agol} E., et~al., 2011, \aj, 142, 72

\bibitem[{Eisenstein} et~al.(2005){Eisenstein}, {Zehavi}, {Hogg}
  et~al.]{eisenstein05a}
{Eisenstein} D.~J., {Zehavi} I., {Hogg} D.~W., et~al., 2005, \apj, 633, 560

\bibitem[{Francis} et~al.(1991){Francis}, {Hewett}, {Foltz}, {Chaffee},
  {Weymann} \& {Morris}]{francis91a}
{Francis} P.~J., {Hewett} P.~C., {Foltz} C.~B., {Chaffee} F.~H., {Weymann}
  R.~J., {Morris} S.~L., 1991, \apj, 373, 465

\bibitem[{Fukugita} et~al.(1996){Fukugita}, {Ichikawa}, {Gunn}, {Doi},
  {Shimasaku} \& {Schneider}]{fukugita96a}
{Fukugita} M., {Ichikawa} T., {Gunn} J.~E., {Doi} M., {Shimasaku} K.,
  {Schneider} D.~P., 1996, \aj, 111, 1748

\bibitem[{Gilbert} \& {Peterson}(2003)]{gilbert03}
{Gilbert} K.~M., {Peterson} B.~M., 2003, \apj, 587, 123

\bibitem[{Green} et~al.(2001){Green}, {Forster} \& {Kuraszkiewicz}]{green01a}
{Green} P.~J., {Forster} K., {Kuraszkiewicz} J., 2001, \apj, 556, 727

\bibitem[{Gunn} et~al.(2006){Gunn}, {Siegmund}, {Mannery} et~al.]{gunn06a}
{Gunn} J.~E., {Siegmund} W.~A., {Mannery} E.~J., et~al., 2006, \aj, 131, 2332

\bibitem[{Harris} et~al.(2016){Harris}, {Jensen}, {Suzuki} et~al.]{harris16a}
{Harris} D.~W., {Jensen} T.~W., {Suzuki} N., et~al., 2016, \aj, 151, 155

\bibitem[{Hewett} et~al.(1995){Hewett}, {Foltz} \& {Chaffee}]{hewett95a}
{Hewett} P.~C., {Foltz} C.~B., {Chaffee} F.~H., 1995, \aj, 109, 1498

\bibitem[{Hutchinson} et~al.(2016){Hutchinson}, {Bolton}, {Dawson}
  et~al.]{hutchinson16a}
{Hutchinson} T.~A., {Bolton} A.~S., {Dawson} K.~S., et~al., 2016,
  arXiv:1607.02432

\bibitem[{Ivashchenko} et~al.(2014){Ivashchenko}, {Sergijenko} \&
  {Torbaniuk}]{ivashchenko14}
{Ivashchenko} G., {Sergijenko} O., {Torbaniuk} O., 2014, \mnras, 437, 3343

\bibitem[{Kinney} et~al.(1990){Kinney}, {Rivolo} \& {Koratkar}]{kinney90a}
{Kinney} A.~L., {Rivolo} A.~R., {Koratkar} A.~P., 1990, \apj, 357, 338

\bibitem[{Kirkpatrick} et~al.(2011){Kirkpatrick}, {Schlegel}, {Ross}
  et~al.]{kirkpatrick11a}
{Kirkpatrick} J.~A., {Schlegel} D.~J., {Ross} N.~P., et~al., 2011, \apj, 743,
  125

\bibitem[{Kollmeier} et~al.(2006){Kollmeier}, {Onken}, {Kochanek}
  et~al.]{kollmeier06a}
{Kollmeier} J.~A., {Onken} C.~A., {Kochanek} C.~S., et~al., 2006, \apj, 648,
  128

\bibitem[{Korista} et~al.(1998){Korista}, {Baldwin} \& {Ferland}]{korista98a}
{Korista} K., {Baldwin} J., {Ferland} G., 1998, \apj, 507, 24

\bibitem[{Lawrence} et~al.(2007){Lawrence}, {Warren}, {Almaini}
  et~al.]{lawrence07a}
{Lawrence} A., {Warren} S.~J., {Almaini} O., et~al., 2007, \mnras, 379, 1599

\bibitem[{Luo} et~al.(2015){Luo}, {Brandt}, {Hall} et~al.]{luo15a}
{Luo} B., {Brandt} W.~N., {Hall} P.~B., et~al., 2015, \apj, 805, 122

\bibitem[{Margala} et~al.(2015){Margala}, {Kirkby}, {Dawson}, {Bailey},
  {Blanton} \& {Schneider}]{margala15a}
{Margala} D., {Kirkby} D., {Dawson} K., {Bailey} S., {Blanton} M., {Schneider}
  D.~P., 2015, arXiv:1506.04790

\bibitem[{Martin} et~al.(2005){Martin}, {Fanson}, {Schiminovich}
  et~al.]{martin05a}
{Martin} D.~C., {Fanson} J., {Schiminovich} D., et~al., 2005, \apjl, 619, L1

\bibitem[{Marziani} et~al.(2003){Marziani}, {Zamanov}, {Sulentic} \&
  {Calvani}]{marziani03}
{Marziani} P., {Zamanov} R.~K., {Sulentic} J.~W., {Calvani} M., 2003, \mnras,
  345, 1133

\bibitem[{Mejia-Restrepo} et~al.(2016){Mejia-Restrepo}, {Trakhtenbrot}, {Lira},
  {Netzer} \& {Capellupo}]{mejia16a}
{Mejia-Restrepo} J.~E., {Trakhtenbrot} B., {Lira} P., {Netzer} H., {Capellupo}
  D.~M., 2016, \mnras

\bibitem[{Merloni}(2016)]{merloni16a}
{Merloni} A., 2016, in { Lecture Notes in Physics, Berlin Springer Verlag\/},
  edited by F.~{Haardt}, V.~{Gorini}, U.~{Moschella}, A.~{Treves}, M.~{Colpi},
  vol. 905 of { Lecture Notes in Physics, Berlin Springer Verlag\/},  101

\bibitem[{Myers} et~al.(2015){Myers}, {Palanque-Delabrouille}, {Prakash}
  et~al.]{myers15a}
{Myers} A.~D., {Palanque-Delabrouille} N., {Prakash} A., et~al., 2015, \apjs,
  221, 27

\bibitem[{Netzer} et~al.(1992){Netzer}, {Laor} \& {Gondhalekar}]{netzer92a}
{Netzer} H., {Laor} A., {Gondhalekar} P.~M., 1992, \mnras, 254, 15

\bibitem[{Netzer} et~al.(2007){Netzer}, {Lira}, {Trakhtenbrot}, {Shemmer} \&
  {Cury}]{netzer07a}
{Netzer} H., {Lira} P., {Trakhtenbrot} B., {Shemmer} O., {Cury} I., 2007, \apj,
  671, 1256

\bibitem[{Noterdaeme} et~al.(2012){Noterdaeme}, {Petitjean}, {Carithers}
  et~al.]{noterdaeme12a}
{Noterdaeme} P., {Petitjean} P., {Carithers} W.~C., et~al., 2012, \aap, 547, L1

\bibitem[{Page} et~al.(2005){Page}, {Reeves}, {O'Brien} \& {Turner}]{page05}
{Page} K.~L., {Reeves} J.~N., {O'Brien} P.~T., {Turner} M.~J.~L., 2005, \mnras,
  364, 195

\bibitem[{P{\^a}ris} et~al.(2011){P{\^a}ris}, {Petitjean}, {Rollinde}
  et~al.]{paris11a}
{P{\^a}ris} I., {Petitjean} P., {Rollinde} E., et~al., 2011, \aap, 530, A50

\bibitem[{P{\^a}ris} et~al.(2016){P{\^a}ris}, {Petitjean}, {Ross}
  et~al.]{paris16a}
{P{\^a}ris} I., {Petitjean} P., {Ross} N.~P., et~al., 2016, arXiv:1608.06483

\bibitem[{Pogge} \& {Peterson}(1992)]{pogge92a}
{Pogge} R.~W., {Peterson} B.~M., 1992, \aj, 103, 1084

\bibitem[{Ricci} et~al.(2013){Ricci}, {Paltani}, {Ueda} \& {Awaki}]{ricci13}
{Ricci} C., {Paltani} S., {Ueda} Y., {Awaki} H., 2013, \mnras, 435, 1840

\bibitem[{Richards} et~al.(2002){Richards}, {Fan}, {Newberg}
  et~al.]{richards02a}
{Richards} G.~T., {Fan} X., {Newberg} H.~J., et~al., 2002, \aj, 123, 2945

\bibitem[{Richards} et~al.(2011){Richards}, {Kruczek}, {Gallagher}
  et~al.]{richards11}
{Richards} G.~T., {Kruczek} N.~E., {Gallagher} S.~C., et~al., 2011, \aj, 141,
  167

\bibitem[{Richards} et~al.(2009){Richards}, {Myers}, {Gray}
  et~al.]{richards09a}
{Richards} G.~T., {Myers} A.~D., {Gray} A.~G., et~al., 2009, \apjs, 180, 67

\bibitem[{Ross} et~al.(2012){Ross}, {Myers}, {Sheldon} et~al.]{ross12a}
{Ross} N.~P., {Myers} A.~D., {Sheldon} E.~S., et~al., 2012, \apjs, 199, 3

\bibitem[{Scott} et~al.(2004){Scott}, {Kriss}, {Brotherton} et~al.]{scott04a}
{Scott} J.~E., {Kriss} G.~A., {Brotherton} M., et~al., 2004, \apj, 615, 135

\bibitem[{Selsing} et~al.(2016){Selsing}, {Fynbo}, {Christensen} \&
  {Krogager}]{selsing16}
{Selsing} J., {Fynbo} J.~P.~U., {Christensen} L., {Krogager} J.-K., 2016, \aap,
  585, A87

\bibitem[{Shakura} \& {Sunyaev}(1973)]{shakura73a}
{Shakura} N.~I., {Sunyaev} R.~A., 1973, \aap, 24, 337

\bibitem[{Shang} et~al.(2004){Shang}, {Brotherton}, {Green} et~al.]{shang04a}
{Shang} Z., {Brotherton} M.~S., {Green} R.~F., et~al., 2004, astro-ph/0409697

\bibitem[{Shang} et~al.(2003){Shang}, {Wills}, {Robinson} et~al.]{shang03}
{Shang} Z., {Wills} B.~J., {Robinson} E.~L., et~al., 2003, \apj, 586, 52

\bibitem[{Shen} et~al.(2015){Shen}, {Brandt}, {Dawson} et~al.]{shen15a}
{Shen} Y., {Brandt} W.~N., {Dawson} K.~S., et~al., 2015, \apjs, 216, 4

\bibitem[{Shen} \& {Ho}(2014)]{shen14}
{Shen} Y., {Ho} L.~C., 2014, \nat, 513, 210

\bibitem[{Shields}(2007)]{shields07a}
{Shields} J.~C., 2007, in { The Central Engine of Active Galactic Nuclei\/},
  edited by L.~C. {Ho}, J.-W. {Wang}, vol. 373 of { Astronomical Society of the
  Pacific Conference Series\/},  355

\bibitem[{Shull} et~al.(2012){Shull}, {Stevans} \& {Danforth}]{shull12}
{Shull} J.~M., {Stevans} M., {Danforth} C.~W., 2012, \apj, 752, 162

\bibitem[{Smee} et~al.(2013){Smee}, {Gunn}, {Uomoto} et~al.]{smee13a}
{Smee} S.~A., {Gunn} J.~E., {Uomoto} A., et~al., 2013, \aj, 146, 32

\bibitem[{Stevans} et~al.(2014){Stevans}, {Shull}, {Danforth} \&
  {Tilton}]{stevans14a}
{Stevans} M.~L., {Shull} J.~M., {Danforth} C.~W., {Tilton} E.~M., 2014, \apj,
  794, 75

\bibitem[{Stoughton} et~al.(2002){Stoughton}, {Lupton}, {Bernardi}
  et~al.]{stoughton02a}
{Stoughton} C., {Lupton} R.~H., {Bernardi} M., et~al., 2002, \aj, 123, 485

\bibitem[{Sulentic} et~al.(2007){Sulentic}, {Bachev}, {Marziani}, {Negrete} \&
  {Dultzin}]{sulentic07a}
{Sulentic} J.~W., {Bachev} R., {Marziani} P., {Negrete} C.~A., {Dultzin} D.,
  2007, \apj, 666, 757

\bibitem[{Sulentic} et~al.(2014){Sulentic}, {Marziani}, {del Olmo}, {Dultzin},
  {Perea} \& {Alenka Negrete}]{sulentic14a}
{Sulentic} J.~W., {Marziani} P., {del Olmo} A., {Dultzin} D., {Perea} J.,
  {Alenka Negrete} C., 2014, \aap, 570, A96

\bibitem[{Sulentic} et~al.(2000){Sulentic}, {Marziani} \&
  {Dultzin-Hacyan}]{sulentic00a}
{Sulentic} J.~W., {Marziani} P., {Dultzin-Hacyan} D., 2000, \araa, 38, 521

\bibitem[{Sulentic} et~al.(2009){Sulentic}, {Marziani} \& {Zamfir}]{sulentic09}
{Sulentic} J.~W., {Marziani} P., {Zamfir} S., 2009, New Astronomy Reviews, 53,
  198

\bibitem[{Suzuki}(2006)]{suzuki06a}
{Suzuki} N., 2006, \apjs, 163, 110

\bibitem[{Telfer} et~al.(2002){Telfer}, {Zheng}, {Kriss} \&
  {Davidsen}]{telfer02}
{Telfer} R.~C., {Zheng} W., {Kriss} G.~A., {Davidsen} A.~F., 2002, \apj, 565,
  773

\bibitem[{Tilton} et~al.(2016){Tilton}, {Stevans}, {Shull} \&
  {Danforth}]{tilton16a}
{Tilton} E.~M., {Stevans} M.~L., {Shull} J.~M., {Danforth} C.~W., 2016, \apj,
  817, 56

\bibitem[{Trakhtenbrot} \& {Netzer}(2012)]{trakhtenbrot12a}
{Trakhtenbrot} B., {Netzer} H., 2012, \mnras, 427, 3081

\bibitem[{Vanden Berk} et~al.(2004){Vanden Berk}, {Yip}, {Connolly}, {Jester}
  \& {Stoughton}]{vandenberk04a}
{Vanden Berk} D., {Yip} C., {Connolly} A., {Jester} S., {Stoughton} C., 2004,
  in { AGN Physics with the Sloan Digital Sky Survey\/}, edited by G.~T.
  {Richards}, P.~B. {Hall}, vol. 311 of { Astronomical Society of the Pacific
  Conference Series\/}, ~21

\bibitem[{Vanden Berk} et~al.(2001){Vanden Berk}, {Richards}, {Bauer}
  et~al.]{vandenberk01a}
{Vanden Berk} D.~E., {Richards} G.~T., {Bauer} A., et~al., 2001, \aj, 122, 549

\bibitem[{Volonteri} et~al.(2016){Volonteri}, {Dubois}, {Pichon} \&
  {Devriendt}]{volunteri16}
{Volonteri} M., {Dubois} Y., {Pichon} C., {Devriendt} J., 2016, \mnras, 460,
  2979

\bibitem[{Warner} et~al.(2004){Warner}, {Hamann} \& {Dietrich}]{warner04}
{Warner} C., {Hamann} F., {Dietrich} M., 2004, \apj, 608, 136

\bibitem[{White} et~al.(2000){White}, {Becker}, {Gregg} et~al.]{white00b}
{White} R.~L., {Becker} R.~H., {Gregg} M.~D., et~al., 2000, \apjs, 126, 133

\bibitem[{Wu} et~al.(2011){Wu}, {Brandt}, {Hall} et~al.]{wu11a}
{Wu} J., {Brandt} W.~N., {Hall} P.~B., et~al., 2011, \apj, 736, 28

\bibitem[{Wu} et~al.(2009){Wu}, {Vanden Berk}, {Brandt}, {Schneider}, {Gibson}
  \& {Wu}]{wu09a}
{Wu} J., {Vanden Berk} D.~E., {Brandt} W.~N., {Schneider} D.~P., {Gibson}
  R.~R., {Wu} J., 2009, \apj, 702, 767

\bibitem[{Xu} et~al.(2008){Xu}, {Bian}, {Yuan} \& {Huang}]{xuy08a}
{Xu} Y., {Bian} W.-H., {Yuan} Q.-R., {Huang} K.-L., 2008, \mnras, 389, 1703

\bibitem[{Y{\`e}che} et~al.(2010){Y{\`e}che}, {Petitjean}, {Rich}
  et~al.]{yeche10a}
{Y{\`e}che} C., {Petitjean} P., {Rich} J., et~al., 2010, \aap, 523, A14

\bibitem[{Yip} et~al.(2004){Yip}, {Connolly}, {Vanden Berk} et~al.]{yip04}
{Yip} C.~W., {Connolly} A.~J., {Vanden Berk} D.~E., et~al., 2004, \aj, 128,
  2603

\bibitem[{York} et~al.(2000){York}, {Adelman}, {Anderson} et~al.]{york00a}
{York} D.~G., {Adelman} J., {Anderson} J.~E., et~al., 2000, \aj, 120, 1579

\bibitem[{Zamorani} et~al.(1992){Zamorani}, {Marano}, {Mignoli}, {Zitelli} \&
  {Boyle}]{zamorani92a}
{Zamorani} G., {Marano} B., {Mignoli} M., {Zitelli} V., {Boyle} B.~J., 1992,
  \mnras, 256, 238

\bibitem[{Zheng} et~al.(1997){Zheng}, {Kriss}, {Telfer}, {Grimes} \&
  {Davidsen}]{zheng97}
{Zheng} W., {Kriss} G.~A., {Telfer} R.~C., {Grimes} J.~P., {Davidsen} A.~F.,
  1997, \apj, 475, 469

\end{thebibliography}

\end{document}